\documentclass[nonacm,sigplan]{acmart}

\AtBeginDocument{%
  }

\usepackage[]{hyperref}

\usepackage{booktabs}
\usepackage{xspace}
\usepackage{mathtools}
\usepackage[ruled, vlined]{algorithm2e}
\usepackage{enumitem}
\usepackage{subcaption}
\usepackage{svg}
\usepackage{graphicx}
\usepackage{bbding}
\usepackage{multirow}

\usepackage{soul}
\usepackage{makecell}
\usepackage{balance}

\newcommand{\red}[1]{{\textcolor{red}{{#1}}}}

\newcommand{\ffc}[1]{{\textcolor{cyan}{{#1}}}}
\newcommand{\old}[1]{{\textcolor{gray}{{#1}}}}
\newcommand{\notsure}[1]{{\textcolor{blue}{{#1}}}}
\newcommand{\revise}[1]{{\textcolor{blue}{{#1}}}}

\newcommand{\name}{Spindle\xspace}
\newcommand{\sname}{Spindle\xspace}

\newcommand{\distmm}{DistMM-MT\xspace}

\newcommand{\mop}{MetaOp\xspace}
\newcommand{\mops}{MetaOps\xspace}
\newcommand{\mlevel}{MetaLevel\xspace}
\newcommand{\mlevels}{MetaLevels\xspace}
\newcommand{\setsub}[1]{\mathcal{S}_{#1}}
\newcommand{\cntnopt}{P_{MPSP}}
\newcommand{\inv}{\text{-1}}

\newtheorem{theorem}{Theorem}

\DeclareMathOperator*{\argmin}{arg\,min}
\DeclareMathSymbol{\myhyphen}{\mathord}{AMSa}{"39}

\newcommand{\wave}{wave\xspace}
\newcommand{\waves}{waves\xspace}

\newif\ifasplos

\ifasplos
\copyrightyear{2025}
\acmYear{2025}
\setcopyright{acmlicensed}\acmConference[ASPLOS '25]{Proceedings of the 30th ACM International Conference on Architectural Support for Programming Languages and Operating Systems, Volume 2}{March 30-April 3, 2025}{Rotterdam, Netherlands}
\acmBooktitle{Proceedings of the 30th ACM International Conference on Architectural Support for Programming Languages and Operating Systems, Volume 2 (ASPLOS '25), March 30-April 3, 2025, Rotterdam, Netherlands}
\acmDOI{10.1145/3676641.3715992}
\acmISBN{979-8-4007-1079-7/2025/03}
\fi

\begin{document}

\title{\sname: Efficient Distributed Training of Multi-Task Large Models via Wavefront Scheduling}


\author{Yujie Wang}
\authornote{School of Computer Science \& Key Lab of High Confidence Software Technologies (MOE), Peking University}
\affiliation{
  \institution{Peking University}
  \city{Beijing}
  \country{China}
}
\email{alfredwang@pku.edu.cn}

\author{Shenhan Zhu}
\authornotemark[1]
\affiliation{
  \institution{Peking University}
  \city{Beijing}
  \country{China}
}
\email{shenhan.zhu@pku.edu.cn}

\author{Fangcheng Fu}
\authornotemark[1]
\affiliation{
  \institution{Peking University}
  \city{Beijing}
  \country{China}
}
\email{ccchengff@pku.edu.cn}

\author{Xupeng Miao}
\affiliation{
  \institution{Purdue University}
  \city{West Lafayette}
  \state{IN}
  \country{USA}
}
\email{xupeng@purdue.edu}

\author{Jie Zhang}
\affiliation{
  \institution{Alibaba Group}
  \city{Beijing}
  \country{China}
}
\email{wanglin.zj@alibaba-inc.com}

\author{Juan Zhu}
\affiliation{
  \institution{Alibaba Group}
  \city{Beijing}
  \country{China}
}
\email{zhujuan.zj@alibaba-inc.com}

\author{Fan Hong}
\affiliation{
  \institution{Alibaba Group}
  \city{Beijing}
  \country{China}
}
\email{hongfan.hf@alibaba-inc.com}

\author{Yong Li}
\affiliation{
  \institution{Alibaba Group}
  \city{Beijing}
  \country{China}
}
\email{jiufeng.ly@alibaba-inc.com}

\author{Bin Cui}
\authornotemark[1]
\authornote{Institute of Computational Social Science, Peking University (Qingdao)}
\affiliation{
  \institution{Peking University}
  \city{Beijing}
  \country{China}
}
\email{bin.cui@pku.edu.cn}

\renewcommand{\shortauthors}{Yujie Wang et al.}
\renewcommand{\shorttitle}{\name: Efficient Distributed Training of Multi-Task \\ Large Models via Wavefront Scheduling}

\begin{abstract}
Recent foundation models are capable of handling multiple tasks and multiple data modalities with the unified base model structure and several specialized model components. However, efficient training of such multi-task (MT) multi-modal (MM) models poses significant system challenges due to the sophisticated model architecture and the heterogeneous workloads of different tasks and modalities.

In this paper, we propose \sname, a brand new training system tailored for resource-efficient and high-performance training of MT MM models via wavefront scheduling.
The key idea of \sname is to decompose the model execution into {\em waves} and address the joint optimization problem sequentially, including both heterogeneity-aware workload parallelization and dependency-driven execution scheduling.
We build our system and evaluate it on various MT MM models. 
Experiments demonstrate the superior performance and efficiency of \name, with speedup ratio up to 71\% compared to state-of-the-art training systems.

\end{abstract}

\maketitle 
\ifasplos
\else
\pagestyle{plain} 
\fi

\section{Introduction}
\label{sec:intro}

In recent years, the field of artificial intelligence (AI) has witnessed a paradigm shift with the advent of large-scale foundation models~\cite{DBLP:conf/naacl/BERT,gpt1,gpt2,DBLP:gpt3,gpt4,DBLP:googleT5,DBLP:llama,DBLP:llama2}.
These models are equipped with extensive intrinsic knowledge, enabling them to be increasingly applied to a broad spectrum of downstream applications, including both the language domain (e.g., ChatGPT~\cite{chatgpt}) and many other data modalities (e.g., images~\cite{DBLP:BEiT,DBLP:ViT,DBLP:CLIP,feng2024docpedia,chen2024internvl-1.5}, speech~\cite{DBLP:wav2vec2.0,DBLP:whisper,DBLP:speech2text2}, video~\cite{DBLP:VideoMAE,DBLP:ViViT}).
The recent extension further involves composite scenarios~\cite{DBLP:Flamingo,DBLP:GATO,DBLP:UNIFIED-IO,DBLP:OFA,DBLP:OFASys,DBLP:Qwen-VL,DBLP:Gemini,DBLP:AnyMAL}, where models are capable of processing and interpreting data across several tasks simultaneously.
However, training these models is highly resource-intensive, requiring substantial GPU computing power.
For example, Meta has announced to release the world-leading multi-modal model, LLaMA-3~\cite{llama-3}, which has over 400B parameters and is trained on more than 48,000 GPUs.

Existing large model training systems are mainly designed for a single model with only one input data modality. Despite the extensive research and engineering efforts aimed at optimizing these systems from multiple perspectives, including distributed communication~\cite{DBLP:megatron,Deepspeed,DBLP:zero++,DBLP:journals/dase/WangLZP23}, memory management~\cite{DBLP:deepspeed-zero,DBLP:zero-offload,DBLP:activation_checkpoint}, and GPU computation~\cite{DBLP:flash_attn,DBLP:flash_attn_2}, their performance is still limited when it comes to handling the increasingly complex requirements of multi-task (MT) multi-modal (MM) models. We identify two unique obstacles when building training systems for MT MM models.

One is the \textit{workload heterogeneity} between different modalities or tasks. On the one hand, MM models often handle data that vary significantly in structure and size, demanding specialized preprocessing and computational approaches. For example, language models (e.g., GPT-family~\cite{gpt1,gpt2,DBLP:gpt3,gpt4}, LLaMA-family~\cite{DBLP:llama,DBLP:llama2}) are usually equipped with dozens of layers with the same configuration (e.g., hidden size), while vision models may involve uneven layers to compute in various resolutions~\cite{DBLP:swin}. On the other hand, multiple tasks may leverage distinct data flows and activate individual model components, leading to inter-task workload heterogeneity. Existing training systems usually overlook such heterogeneity and apply sub-optimal training methodologies.

\begin{figure}[!t]
    \centering
    \includegraphics[width=\linewidth]{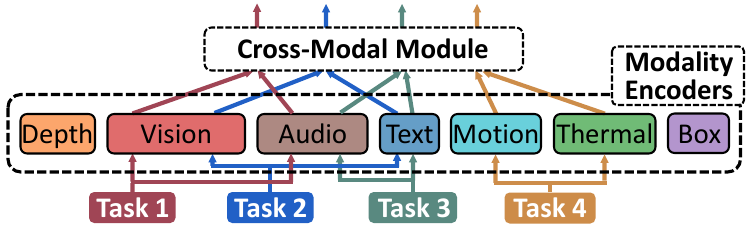}
    \includegraphics[width=\linewidth]{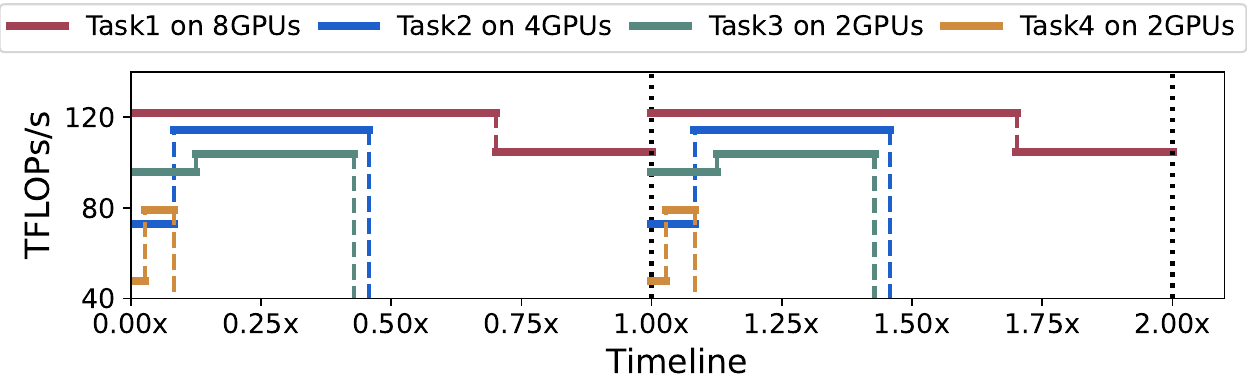}
    
    \caption{
    The upper portion illustrates the general model structure and training flow of MT MM training. The lower portion displays the current device utilization, measured in FLOPs per second, during the decoupled execution of four tasks across 2 iterations. 
    Utilization fluctuation of different-colored and same-colored lines indicate inter-task and intra-task workload heterogeneity, respectively.
    }
    \label{fig:intro_observation}
\end{figure}

Another is the \textit{execution dependency} among different model components. Recent MT MM model development usually adopts a sub-model sharing approach~\cite{DBLP:ImageBind,DBLP:OFA,DBLP:OFASys,DBLP:Qwen-VL,DBLP:AnyMAL}, where partial model layers containing common knowledge are shared across different modalities and tasks. As shown in Fig.~\ref{fig:intro_observation}, each data type also has its own learning component. Within every training iteration, the input data mixed with multiple modalities is simultaneously fed into the sophisticated model, where different model components are intricately activated and updated. 
To avoid redundant resource usage, the shared components are usually responsible for the data flows from multiple sources, resulting in execution barriers and blocking the following model layers.
In addition, the proportion of different data modalities in MT workloads may shift over time due to task addition and completion, introducing further training complexity.
To the best of our knowledge, none of existing training systems can deal with these unforeseen dependency efficiently due to the lack of understanding MT MM model execution.

To address these challenges, this paper introduces \sname, a resource-efficient and high-performance training system for large-scale MT MM models.
Considering the workload heterogeneity and execution dependency, a na{\"i}ve solution is to \textit{decouple} the model structure based on modality and task, replicate the shared components, and deploy them on separate devices. In this way, each sub-model can be optimized by existing systems, but it also brings significant \textit{resource wastage and underutilization}, as well as additional overheads from replica synchronization.
As an example, Fig.~\ref{fig:intro_observation} showcases that such a na{\"i}ve, decoupled execution suffers from fluctuating device utilization both intra-task and inter-task due to workload heterogeneity, leading to low or even idle GPU utilization for some time slots.

Instead of decoupling, \sname manages to directly train the whole complex model \textit{without} disjoint sub-model to minimize the resource usage.
A key insight behind \sname's design is that \textit{heterogeneous} and \textit{dependent} sub-models can be decomposed into several sequentially executed \textit{waves}.
Specifically, \name treats a \textit{\wave} as the smallest scheduling unit for execution.
Within each \wave, the runtime engine concurrently executes multiple sliced \textit{\mops} (i.e., continuously identical operators, as defined in \S\ref{subsec:contraction}).
These sliced \mops are distributed across distinct and fixed groups of devices, while ensuring that their execution time costs are balanced.
The concept of \wave is central to \name's design and operation.
A more comprehensive definition of \wave will be provided in \S\ref{subsec:scheduler}, along with a concrete example of 6 \waves illustrated in Fig.~\ref{fig:mtp_execution_plan}.

To achieve resource-efficient and high-performance training of MT MM models, there are three key challenges for \sname to address. In the following, we will introduce each challenge and how \sname solves them.

First, finding the optimal model parallel configuration for heterogeneous workloads with diverse computational characteristics is a complex combinatorial problem. Existing single-model automatic parallelization approaches (e.g., Alpa~\cite{DBLP:ALPA}, Unity~\cite{unger2022unity}, Galvatron~\cite{DBLP:Galvatron,Galvatron-BMW}) assume a spatial pipeline stage partition, and each operator (Op) is executed by all devices of the corresponding pipeline stage. Unfortunately, such assumptions only work for homogeneous models, failing to adapt to heterogeneous MT MM models.

Instead of solving the parallel configuration directly, \sname captures the workload heterogeneity at the operator granularity and estimates its execution overheads under different amount of allocated resources and parallel configurations (\S\ref{subsec:scale_est}). The final configuration decision is left to the later step since it requires to be jointly optimized with considering the execution dependency. \sname also introduces \mop to contract the graph (i.e., fusing continuous identical operators) to avoid redundant estimation overheads and shrink the problem scale (\S\ref{subsec:contraction}).

Second, breaking down the whole model into sequentially executed waves may easily result in inefficiencies.
Determining the optimal division of waves is complicated since the operators differ significantly in execution overheads and have intricate operator dependencies.

\sname addresses this problem with two steps: 1) \sname's \textit{resource allocator} (\S\ref{subsec:allocator}) traverses the computation graph following the dependency topology and decides the optimal resource allocation for \mops in each candidate set (i.e., currently executable \mops). Here we reformulate this issue as a malleable project scheduling problem (MPSP) and subsequently derive the optimal solution. 2) After obtaining the parallel configuration of each \mop, the \textit{wavefront scheduler} (\S\ref{subsec:scheduler}) greedily slices and selects \mops to craft compact \waves and minimizes the overall execution time.

Third, given the resource allocation plan and wavefront execution schedule, how to map them into physical devices is still a problem, 
as different mapping may lead to distinct inter-\wave communication overheads and per-device memory consumption.
To further improve the system efficiency, \sname carefully considers these trade-offs and the real environment constraints (e.g., inter-device bandwidth, memory capacity) when generating the device placement plan (\S\ref{subsec:device_place}).

Our contributions are summarized as follows:

\begin{itemize}
    \item We present \sname, a resource-efficient and high-\\performance training system for MT MM models.
    \item We propose a jointly optimization framework to achieve heterogeneity-aware workload parallelization and \\dependency-driven execution scheduling.
    \item We build a general runtime engine to perform the wavefront schedule, automatically resolving execution dependencies at the wave boundaries.
    \item We evaluate \sname on various MT MM models, and the results demonstrate the superior performance and efficiency of \name compared with the state-of-the-art baselines, with the speedup ratio up to 71\%.
\end{itemize}

\section{Preliminary} \label{sec:preliminary}

\subsection{Multi-Task Multi-Modal Models} \label{sec:pre_models}


Foundation models, such as GPT series~\cite{gpt2,DBLP:gpt3,gpt4}, LLaMA series~\cite{DBLP:llama,DBLP:llama2}, have set new benchmarks across various language tasks and revolutionized deep learning.
They've also been successfully adapted for other modalities and tasks, including image processing~\cite{DBLP:BEiT,DBLP:ViT,DBLP:swin}, audio processing~\cite{DBLP:wav2vec2.0,DBLP:whisper,DBLP:speech2text2}, video analysis~\cite{DBLP:VideoMAE,DBLP:ViViT}.
Multi-modal models~\cite{DBLP:CLIP,DBLP:ImageBind,DBLP:OFA,DBLP:OFASys,DBLP:Qwen-VL,DBLP:Qwen-Audio} leverage these foundation models to integrate information from multiple data modalities.
Multi-modal models typically have the multi-tower structure, utilizing multiple modality encoders to extract modality features, and a cross-modal module for feature alignment and fusing.
Some of these models fuse modality information via lightweight contrastive learning objectives~\cite{DBLP:CLIP,DBLP:ALIGN,DBLP:Florence,DBLP:VideoCLIP,DBLP:Audioclip, DBLP:ImageBind,DBLP:journals/dase/XuLLY24}, with CLIP~\cite{DBLP:CLIP} being a notable example, and ImageBind~\cite{DBLP:ImageBind} further extending CLIP to six modalities. 
Others fuse modalities via the language model with generative loss~\cite{DBLP:OFA,DBLP:BLIP,DBLP:CoCa,DBLP:BEIT3,DBLP:BLIP-2,DBLP:PaLM-E,DBLP:MiniGPT-4,DBLP:LLAVA}, some leveraging the powerful pretrained LLMs.

Recently, researchers have begun to construct more complicated multi-task multi-modal (MT MM) models~\cite{DBLP:Flamingo,DBLP:OFASys,DBLP:Qwen-VL,DBLP:AnyMAL,DBLP:journals/jcst/QinLTHGP24}, enabling processing diverse multi-modal tasks within a unified model.
This is because each modality encompasses various tasks, and each task often involves multiple modalities as well, and this reflects researchers' aspiration towards general-purpose AI.
Fig.~\ref{fig:intro_observation} (upper side) illustrates the general structure and training flow of MT MM models.
Flamingo~\cite{DBLP:Flamingo} is among the first to handle multiple vision-language tasks.
OFASys~\cite{DBLP:OFASys} proposes a general MT MM learning paradigm,
as shown in Fig.~\ref{fig:intro_observation}, 
designing distinct modality encoders and cross-modal modules for different tasks and modalities, allowing the activation of different components as required by the task and modality at hand. 
For example, speech recognition and image captioning tasks shall activate and share the text encoder but feed the visual- and audio-inputs into different encoders.
Many empirical results~\cite{DBLP:Flamingo,DBLP:GATO,DBLP:UNIFIED-IO,DBLP:OFA,DBLP:OFASys,DBLP:Qwen-VL,DBLP:journals/jcst/QinLTHGP24} have also shown that such a joint multi-task training paradigm achieves better multi-modal capabilities for MT MM models than performing single-task training separately.

\subsection{Parallelisms in Distributed Training}
As model sizes and training data volumes grow, modern DL systems commonly employ various parallelism techniques for distributed training on GPU clusters.
Data parallelism (DP)~\cite{DBLP:pytorch-ddp,DBLP:deepspeed-zero,DBLP:pytorch-fsdp} splits the input data, with each device handling a portion of the data storage and computation, and synchronizing model gradients across devices. 
Model parallelism~\cite{DBLP:megatron,DBLP:conf/nips/gpipe,DBLP:conf/sosp/pipedream,pipedream-flush,DBLP:journals/jcst/GuanLLWGL24} partitions model parameters, with each device responsible for a segment of the model. Model parallelisms can be categorized into two popular types: tensor parallelism (TP) partitions the model vertically~\cite{DBLP:megatron}, while pipeline parallelism (PP)~\cite{DBLP:conf/nips/gpipe,DBLP:conf/sosp/pipedream,pipedream-flush} splits the model horizontally, organizing model execution into a pipeline.
Contemporary distributed training systems, such as Megatron-LM~\cite{DBLP:megatron} and DeepSpeed~\cite{Deepspeed}, leverage multiple parallelisms and implement a hybrid parallelism approach for model training. For example, Megatron-LM introduces 3D parallelism, which concurrently utilizes DP, TP, and PP.
Researchers have also developed advanced automatic parallelism~\cite{DBLP:conf/mlsys/flexflow,DBLP:ALPA,DBLP:Galvatron,Galvatron-BMW} techniques to facilitate the tuning of optimal parallelism combinations, which integrates multiple parallelism dimensions, employ sophisticated optimization workflows, and automatically determine the most efficient hybrid parallelism strategy.
However, these existing training system are mainly designed for single task and single model training, with limited performance on the complex scenario of training MT MM models.

\newcommand{\comp}{{(\text{c})}}
\newcommand{\redu}{{(\text{r})}}
\newcommand{\ptp}{{(\text{p})}}
\newcommand{\node}{{(\text{node})}}
\newcommand{\DP}{{(\text{d})}}
\newcommand{\TP}{{(\text{t})}}
\newcommand{\send}{{(\text{s})}}
\newcommand{\recv}{{(\text{r})}}
\newcommand{\dscr}{{(\text{d})}}
\newcommand{\bfa}{\mathbf{a}}


\begin{figure}[!t]
    \centering
    \includegraphics[width=\linewidth]{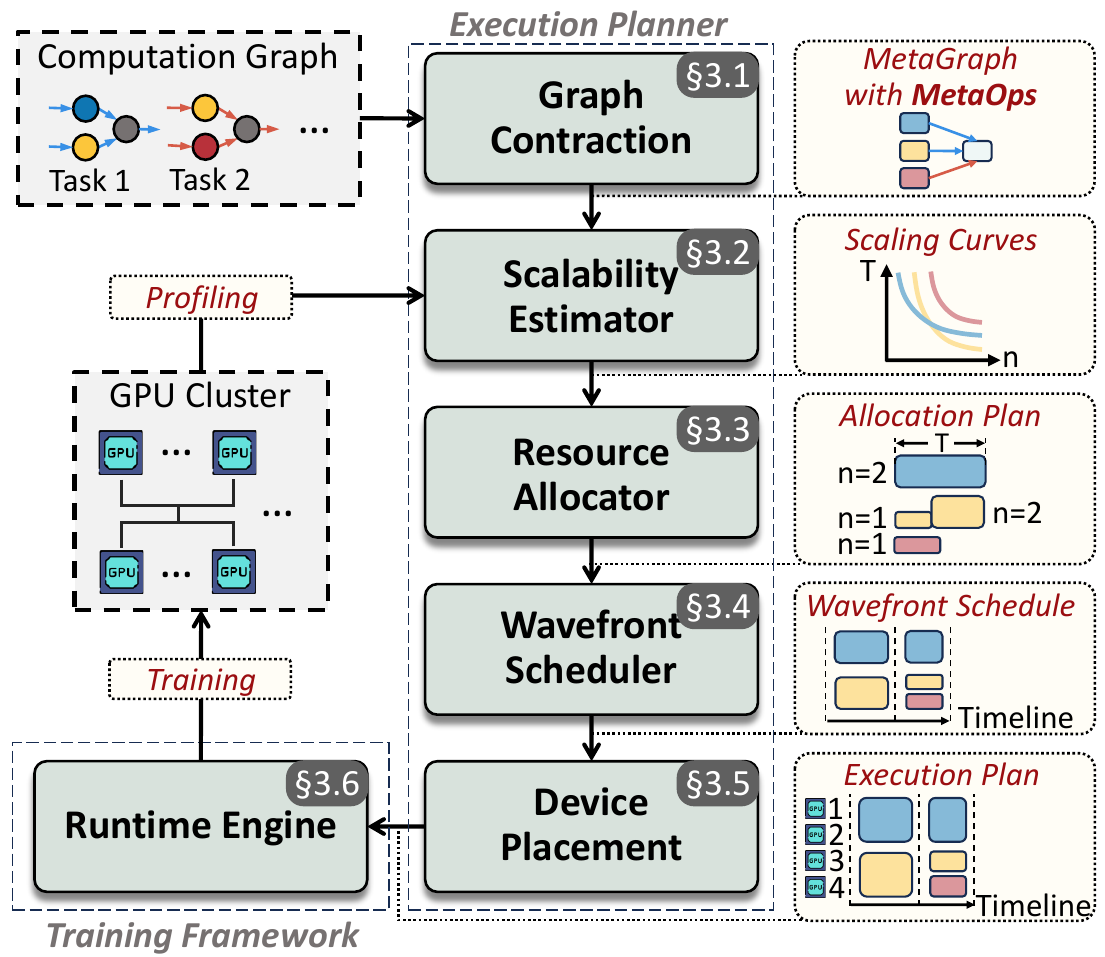}
    \caption{
    Architecture overview of \sname.
    }
    \label{fig:mtp_overview}
\end{figure}

\section{System Design} 
\label{sec:system_design}

\name is a highly efficient and scalable training framework for MT MM models. 
%
Fig.~\ref{fig:mtp_overview} depicts its system architecture, comprising the execution planner and the training framework. 
Given the diverse user-defined training tasks and the GPU cluster, the goal of \name is to devise the most efficient execution plan to facilitate effective MT MM training.

\subsubsection*{Problem Formulation}
We formalize the optimization problem of \name as follows.
Firstly, \name interprets the input tasks as a unified directed acyclic computation graph $\mathcal{G} = (\mathcal{V}, \mathcal{E})$, where each node $i \in \mathcal{V}$ represents a computational operator and each edge $\langle i,j \rangle \in \mathcal{E}$ denotes the data flow from operator $i$ to $j$. 
Each task activates specific operators and parameters with unique data flows. 
For instance, a vision-related task activates a vision Transformer layer as an operator, with image features serving as the data flow.
The left side of Fig.~\ref{fig:graph_contraction} displays an example of a computation graph.
Then, given the computation graph $\mathcal{G}$ and the GPU cluster with $N$ devices, \name aims to minimize the maximal operator completion time $C$.
Specifically, we need to find an execution plan $P$, which assigns each operator $i \in \mathcal{V}$ with an \textbf{AS}-tuple $\langle n_i, s_i \rangle \in\mathcal{U}$, such that the operator $i$ is \textbf{A}llocated $n_i$ devices and is \textbf{S}cheduled to execute from time $s_i$.
Here the set $\mathcal{U} = \{\langle n, s \rangle | n \in \mathbb{N}, s \geq 0\}$ is formed by all valid AS-tuples. We further denote the execution time of operator $i$ when allocated $n_i$ devices as $t_i=T_i(n_i)$.
Then, the optimization problem is formulated as follows. 
Here~\eqref{eq:ori_alloc_capacity} is the allocation capacity constraint for any time $t$, and~\eqref{eq:ori_op_dep} is the operator dependency constraint.
\begin{align}
    & \argmin_{\substack{P=\{i \rightarrow \langle n_i,s_i \rangle |\\ i \in \mathcal{V}, \langle n_i,s_i \rangle \in \mathcal{U}\}}} C \coloneqq \max_{i \in \mathcal{V}}\{s_i + t_i\} 
    \label{eq:ori_formulation} \\
    \text{s.t.} 
    & \sum_{{t \in (s_i, s_i+t_i), i \in \mathcal{V}}} n_i \leq N \;\;\; \text{for }\forall t \in \mathbb{R}^+ \label{eq:ori_alloc_capacity}\\
    & s_i+t_i \leq s_j \;\;\; \text{for }\forall \langle i,j \rangle \in \mathcal{E} \label{eq:ori_op_dep}
\end{align}

\subsubsection*{Sketch of Solution}
Before stepping into the solution, we'd like to first present an overview for better readability.
First, \name initiates a graph contraction process (\S\ref{subsec:contraction}), contracting the original graph $\mathcal{G}$ into a MetaGraph $\mathcal{G}_M$ composed of \textsl{\mops} (Fig.~\ref{fig:graph_contraction}), where each \mop characterizes a unique workload. 
This process further decouples \mops into different \textsl{\mlevels}, ensuring that there are no dependencies among \mops within the same \mlevel.
Second, the scalability estimator (\S\ref{subsec:scale_est}) estimates the execution time and resource scalability for each \mop, producing scaling curves (Fig.~\ref{fig:resrc_scale}). 
Following this, the resource allocator (\S\ref{subsec:allocator}) deduces the allocation plan for each \mlevel individually (Fig.~\ref{fig:mtp_allocator}). 
Given the allocation plan, the wavefront scheduler (\S\ref{subsec:scheduler}) slices the \mops and organizes them into \textsl{\waves}, and produces the wavefront schedule for execution. 
Subsequently, device placement (\S\ref{subsec:device_place}) strategies are then employed to assign \mops to appropriate devices, resulting in the \name execution plan (Fig.~\ref{fig:mtp_execution_plan}).
Finally, the runtime engine (\S\ref{subsec:engine}) utilizes this plan to instantiate the model on each device and facilitate an efficient MT MM training process.


\subsection{Graph Contraction} \label{subsec:contraction}


\subsubsection*{Depicting Workload Heterogeneity with \textsl{\mops}}
\begin{figure}[!t]
    \centering
    \includegraphics[width=\linewidth]{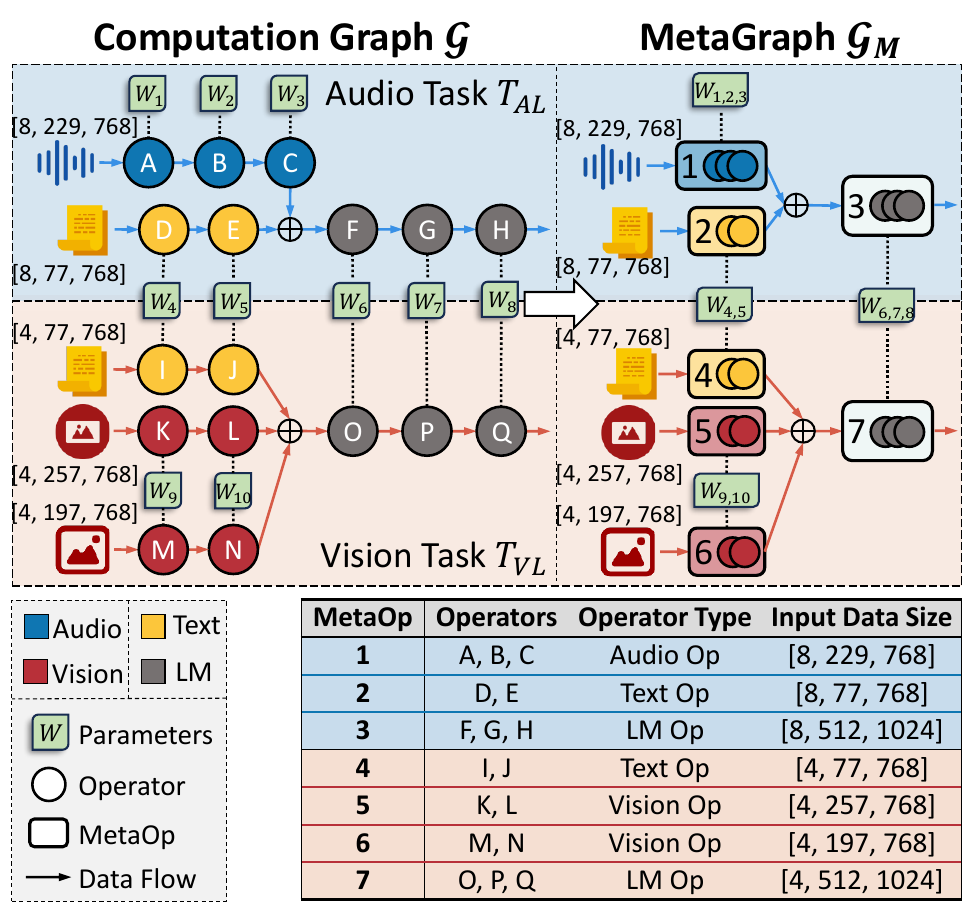}
    \caption{
    Computation graph $\mathcal{G}$ and contracted MetaGraph $\mathcal{G}_M$.
    }
    \label{fig:graph_contraction}
\end{figure}

\name minimizes the execution time by optimizing resource allocation and scheduling for each operator within $\mathcal{G}$. 
This optimization process necessitates an understanding of the workload characteristics for each operator $i \in \mathcal{V}$, which can be reflected by its execution time function $t_i = T_i(n_i)$, which varies with the device allocation amount $n_i$. 
Given that $\mathcal{G}$ typically includes a large number of operators while many of them share similar workload characteristics (such as stacked Transformer layers), \name initiates a graph contraction process to streamline the complicated graph. It categorizes operators based on their computational workload characteristics, as illustrated in Fig.~\ref{fig:graph_contraction}.
In this process, operators are contracted into a \mop if they meet the following criteria:
\begin{enumerate}[label=(\arabic*), nosep]
\item There is a data flow between operator $i$ and $j$, i.e.,  $\langle i,j \rangle \in \mathcal{E}$, and both the out-degree of operator $i$ and the in-degree of operator $j$ are 1, ensuring that they are direct predecessors and successors to each other.
\item Operator $i$ and $j$ share the same operator type and input data size, confirming identical workloads.
\end{enumerate}
%
During graph contraction, 
we traverse the original graph $\mathcal{G}$ in topological order, contracting operators based on the specified criteria until no further pairs of operators meeting these conditions.
This results in a contracted MetaGraph $\mathcal{G}_M = (\mathcal{V}_M, \mathcal{E}_M)$, with each node $m \in \mathcal{V}_M$ representing a \mop that consists of $L_m$ consecutive operators in $\mathcal{G}$. 
Since operators in the same \mop share the same workload, we slightly abuse the notation and denote the execution time function for each operator in \mop $m$ as $T_m(n)$.

\subsubsection*{Disentangling \mop Dependency with \textsl{\mlevels}}
To facilitate operator-level resource allocation and scheduling, we further introduce an abstraction called \mlevel, which signifies the level of dependency. \mops at the same level are independent to each other.
The level of each \mop can be derived by a Breadth-First-Search (BFS), with the level assigned based on the search depth, which inherently ensures no dependency among the \mops of same level.
By doing so, the problem~\eqref{eq:ori_formulation} can be dissected into several simplified sub-problems for different \mlevels. 
Next, we introduce how \name derives the allocation and scheduling for each \mlevel individually, and merges them into the final plan. 

\subsection{Scalability Estimator} \label{subsec:scale_est}

\begin{figure}[!t]
    \centering
    \includegraphics[width=\linewidth]{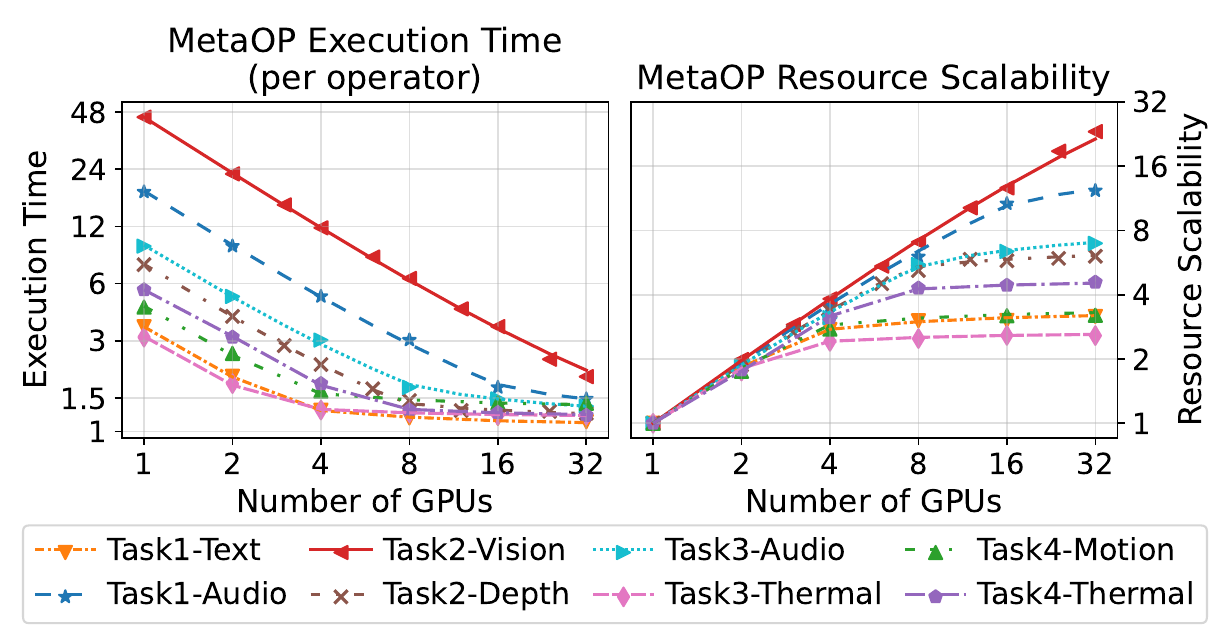}
    \caption{
    An example of the execution time and resource scalability of \mops in 4-task Multitask-CLIP, denoted as \textsl{scaling curves}.
    }
    \label{fig:resrc_scale}
\end{figure}

As {\mops} differ in operator types and/or input data sizes, they characterize heterogeneous workloads and thus necessitate different amount of resources. 
Furthermore, these {\mops} have distinct resource scalability (i.e., how its execution time varies w.r.t. the amount of allocated resources).
For instance, the left side of Fig.~\ref{fig:resrc_scale} shows the execution time of different \mops, $T_m(n)$, in Multitask-CLIP (a multi-task extension of CLIP, detailed in~\S\ref{subsec:exp_setup}). 
Some \mops show almost linear decreases in execution time as resources increase (e.g., Task2-Vision), while others decrease much more slowly (e.g., Task1-Text).
The right side of Fig.~\ref{fig:resrc_scale} further shows the value of $\varsigma_m(n)=T_m(1)/T_m(n)$, which measures how much the operator accelerates when using more GPUs, and a value of $\varsigma_m(n)$ closer to $n$ signifies better resource scalability.
As can be seen, different \mops not only have varying execution time, but also exhibit different resource scalability, posing a significant challenge for resource allocation.

In response to this issue, \name employs a scalability estimator to accurately capture the execution time and the resource scalability of each \mop. 
Previous works~\cite{DBLP:ALPA,unger2022unity,DBLP:Galvatron} have designed effective estimation methods for distributed training, commonly utilizing the $\alpha\myhyphen\beta$ modelling~\cite{DBLP:alpha_beta_model}.
However, although this may work well for homogeneous workloads (e.g., LLMs with homogeneous layers), we find that it does not fit the workload heterogeneous nature of MT MM models. 
This is because different \mops have distinct workload and resource scalability, and the invoked kernels may vary across different per-device workloads, therefore causing distinct performance.
In a nutshell, our scalability estimator adopts the \textit{piecewise} $\alpha\myhyphen\beta$ modelling for more accurate estimation of heterogeneous MT MM workloads.
Given the target MT MM model, it profiles several discrete data points $(n_i, T_m(n_i))$ for each \mop under different parallel configurations, and then fits the curve of piecewise $\alpha\myhyphen\beta$ function. 
To estimate the execution time $T_m(n)$, it locates the range that $n$ falls into, and returns the estimated time according to the corresponding piecewise function. 
In practice, the profiling and estimating process for each MT MM model takes within 5 minutes, which is negligible compared to the training time.
In Fig.~\ref{fig:resrc_scale}, the scatter points represent empirical measurements, while the curves depict the function estimated by our scalability estimator, which we denote as \textsl{scaling curves}. 
As can be seen, our scalability estimator effectively and accurately estimates the execution time $T_m(n)$ for each \mop. 
More details are illustrated in \ifasplos\else{Appendix~\ref{sec:append_estimator}.}\fi

\begin{figure*}[tbp]
    \centering
    \begin{subfigure}[b]{0.4\textwidth}
        \includegraphics[width=\textwidth]{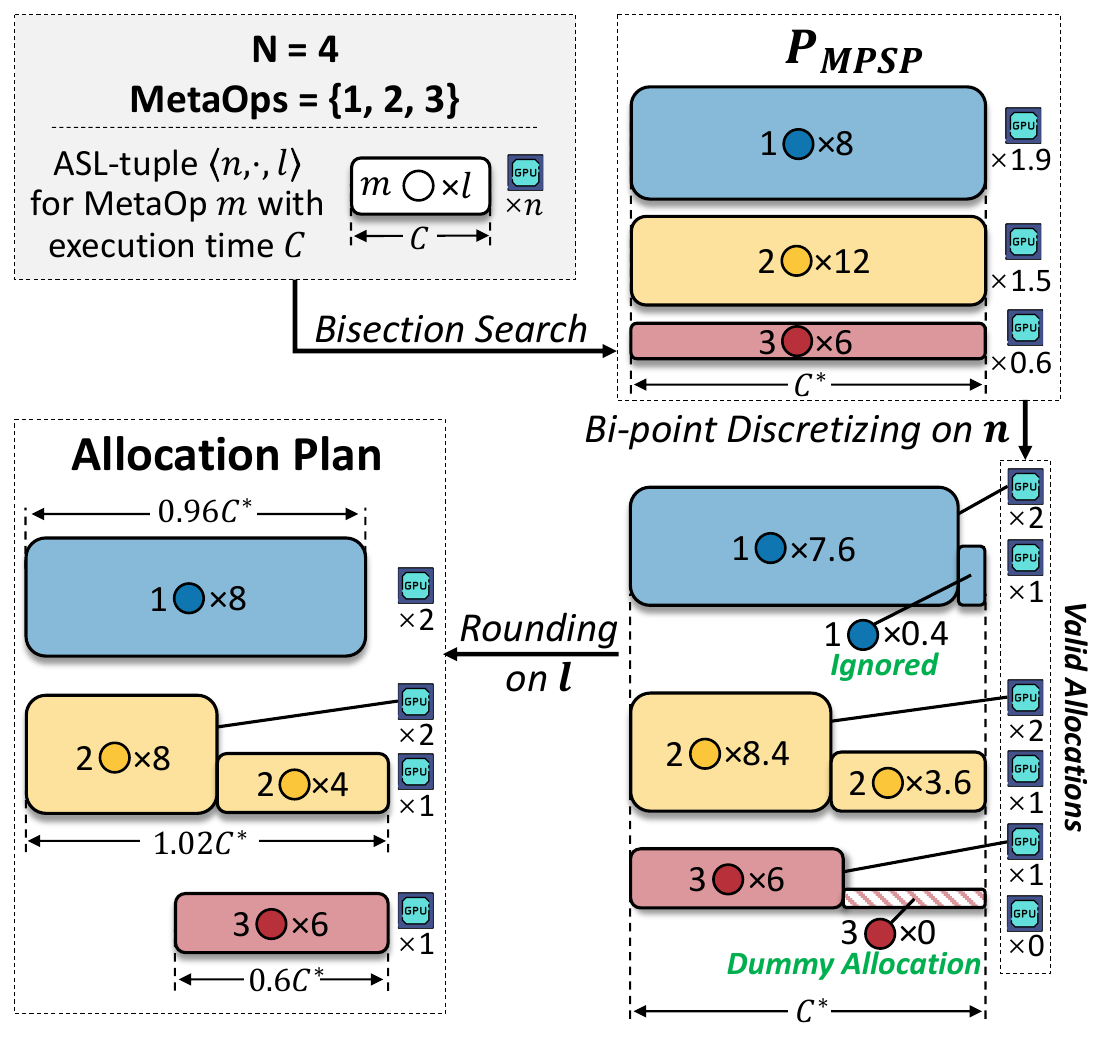}
        \caption{Illustration of workflow of \sname allocator, which allocates resources to $3$ \mops on $4$ devices.}
        \label{fig:mtp_allocator}
    \end{subfigure}
    \begin{subfigure}[b]{0.59\textwidth}
        \includegraphics[width=\textwidth]{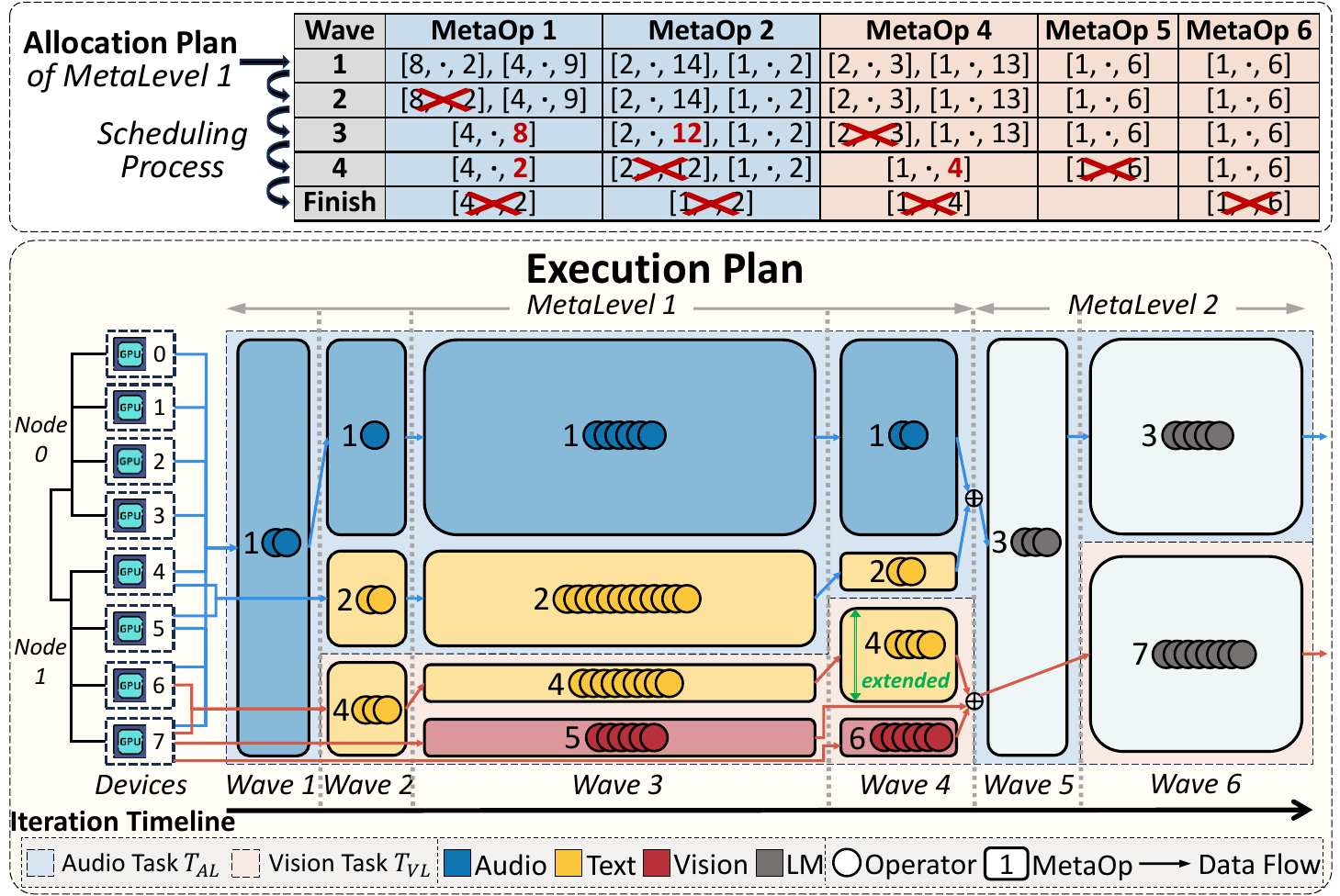}
        \caption{Example of \name execution plan consisting of $6$ \waves.}
        \label{fig:mtp_execution_plan}
    \end{subfigure}
    \caption{
        Illustration of \name allocator and \name execution plan.
    }
\end{figure*}

\subsection{Resource Allocator} \label{subsec:allocator}

We now introduce our resource allocator, which allocates appropriate computational resources to \mops. 
We first transition problem~\eqref{eq:ori_formulation} into the sub-problem on \mlevel. 
We then detail our allocation strategies, which first relax constraints and optimize the continuous problem, and then discretize the optimal solution for practical allocation plans.


\subsubsection*{Problem Formulation on \mlevel}
We first re-formulate the problem~\eqref{eq:ori_formulation} on one \mlevel with a set of \mops denoted by $\mathcal{{\widetilde{V}}}_M$.
In this formulation, we split each \mop into different execution part, by assigning it with several ASL-tuples $\langle n,s,l \rangle \in \mathcal{U}_M$, such that $l$ consecutive operators of this \mop are scheduled to execute from time $s$ with $n$ devices.
Here $\mathcal{U}_M = \{\langle n, s, l \rangle | n,l \in \mathbb{N}, s \geq 0\}$ is formed by all valid ASL-tuples.
For each \mop $m \in \mathcal{\widetilde{V}}_M$, its execution plan is a set of ASL-tuples $P_m$. 
For a \mlevel, the execution plan $P$ consists of $P_m$ for all \mops $m \in \mathcal{\widetilde{V}}_M$, i.e., $P=\{m \rightarrow P_m\}$.
Given $m \in \mathcal{\widetilde{V}}_M$ and one ASL-tuple $p=\langle n_{m}^{(p)},s_{m}^{(p)},l_{m}^{(p)} \rangle \in P_m$, we denote the execution time span, end time, and time interval by 
$t_{m}^{(p)} = T_m(n_{m}^{(p)}) \cdot l_{m}^{(p)}$, 
$e_{m}^{(p)} = s_{m}^{(p)}+t_{m}^{(p)}$, 
and $I_{m}^{(p)}=(s_{m}^{(p)},e_{m}^{(p)})$, 
respectively.
The problem can be re-written as:
{
\begin{align}
    & \argmin_{P=\{m \rightarrow P_m | m \in \mathcal{\widetilde{V}}_M, P_m \subset 2^{\mathcal{U}_M}\}} \widetilde{C} \coloneqq \max_{m \in \mathcal{\widetilde{V}}_M, p \in P_m}\{e_{m}^{(p)}\} \label{eq:meta_problem} \\
    \text{s.t. }
    & \sum_{{t\in I_{m}^{(p)}, m \in \mathcal{\widetilde{V}}_M, p \in P_m}} n_{m}^{(p)} \leq N \;\;\; \text{for }\forall t \in \mathbb{R}^+ \label{eq:meta_alloc_capacity} \\
    & I_{m}^{(p_1)} \cap I_{m}^{(p_2)} = \varnothing \;\;\; \text{for}\;\forall m\in \mathcal{\widetilde{V}}_M, p_1,p_2\in P_m \label{eq:meta_intraop_dep} \\
    & \sum_{p \in P_m} l_{m}^{(p)} = L_m \;\;\; \text{for }\forall m \in \mathcal{\widetilde{V}}_M \label{eq:meta_work_complete}
\end{align}
}
Compared with the original problem~\eqref{eq:ori_formulation}, the sub-problem~\eqref{eq:meta_problem} on \mlevel gets rid of the dependency constraint, while the  constraint~\eqref{eq:meta_intraop_dep} enforces the execution intervals of ASL-tuples in $P_m$ to be pairwise disjoint, because operators within the same \mop cannot execute simultaneously, and~\eqref{eq:meta_work_complete} ensures all operators are executed for each \mop.


\subsubsection*{Optimum of the Continuous Problem}
If we relax the constraints, allowing GPU resources and operators to be continuously divisible 
(i.e., $n$ and $l$ in ASL-tuples are not limited to integers), the problem is transformed into a well-established problem, malleable project scheduling problem (MPSP), with malleable projects and continuously divisible resources~\cite{parallel_scheduling_maciej2009}.
We denote the optimal solution of this relaxed problem by $\cntnopt$.
Prior works~\cite{pcntn_theory_1981, pcntn_theory_1982} have given the following theorem.
\begin{theorem}\label{thm:pcntn}
    If the execution time functions $T_m(n)$, $n\in \mathbb{R}^+$, are positive and non-increasing for every \mop $m\in\mathcal{\widetilde{V}}_M$, then $\cntnopt = \{m \rightarrow P_m\}$ satisfies that $P_m = \{\langle n_m^\ast, 0, L_m\rangle\}, \forall m \in \mathcal{\widetilde{V}}_M$, where the optimum objective $\widetilde{C}^\ast$ and allocations $n_m^\ast$ can be found from
    {
    \begin{equation}
        T_m(n_m^\ast) \cdot L_m = \widetilde{C}^\ast 
        \text{ for }\forall m \in \mathcal{\widetilde{V}}_M
        \text{ and }
        \sum_{m \in \mathcal{\widetilde{V}}_M} n_m^\ast = N.
    \end{equation}
    }
\end{theorem}
From Theorem~\ref{thm:pcntn}, it follows that in the optimal situation, all \mops start simultaneously, execute all their operators, and finish together.
They share an identical end time $e_{m} = \widetilde{C}^\ast$, which is exactly the minimized operator completion time.




To achieve $\cntnopt$, our allocator utilizes the scaling curves from \S\ref{subsec:scale_est} to acquire an estimation of $T_m(n)$, and performs a bisection search procedure over $\widetilde{C}^\ast$ with the following equation. The details are illustrated in \ifasplos\else{Appendix~\ref{sec:append_p_cntn}.}\fi
{
\small
\begin{equation}
    \sum_{m\in \mathcal{\widetilde{V}}_M} T_m^\inv\left({\widetilde{C}^\ast}/{L_m}\right) = N.
\end{equation}
}


\subsubsection*{Bi-point Discretized Allocation}
From the continuous problem, we've determined the optimal time $\widetilde{C}^\ast$, as well as the optimal allocations for each \mop, $n_m^\ast$, which is a real number.
To reinstate $n$'s as integers, our allocator computes each \mop's proper discrete allocations individually.
For every \mop $m$, it uses two discrete ASL-tuples $\langle\overline{n_m}, \cdot, \overline{l_m}\rangle, \langle \underline{n_m}, \cdot, \underline{l_m}\rangle$ to linearly represent the continuous, optimal solution $\langle n_m^\ast, 0, L_m \rangle$ in $\cntnopt$. To preserve the optimum property of $\cntnopt$, we require the discretized allocation plan to satisfy the following two conditions:

\begin{subequations}
\noindent
\begin{minipage}{0.35\linewidth}
\begin{equation}\label{eq:bp_dscr_alloc_1}
\centering
    L_m = \overline{l_m} + \underline{l_m}
\end{equation}
\end{minipage}%
\begin{minipage}{0.03\linewidth}
\centering
$ $
\end{minipage}%
\begin{minipage}{0.61\linewidth}
\begin{equation}\label{eq:bp_dscr_alloc_2}
\centering
    \widetilde{C}^\ast = T_m(\overline{n_m}) \cdot \overline{l_m} + T_m(\underline{n_m}) \cdot \underline{l_m}
\end{equation}
\end{minipage}{\vskip1em}
\end{subequations}
\noindent
Cond.~\eqref{eq:bp_dscr_alloc_1} ensures these two discrete ASL-tuples complete the workload of \mop $m$, and Cond.~\eqref{eq:bp_dscr_alloc_2} ensures their total execution time is exactly equal to the minimum operator completion time $\widetilde{C}^\ast$ in $\cntnopt$, thus perserving the optimum property.
Here we first select $\overline{n_m}, \underline{n_m}$ as the closest \textsl{valid} integer numbers such that $n_m^\ast \in [\underline{n_m}, \overline{n_m}]$, and $\overline{l_m},\underline{l_m} \in \mathbb{R}^+$ are derived naturally.
For instance, as shown in Fig.~\ref{fig:mtp_allocator}, \mop $2$ with $n_2^\ast=1.5, L_2=12$ in $\cntnopt$ is discretized as $\overline{n_2}=2, \underline{n_2}=1$ and $\overline{l_2}=8.4, \underline{l_2}=3.6$ in this step.
Here we impose the \textsl{valid} constraint on the allocation $n$ for \mop $m$ for practical reasons.
For instance, if an \mop is applied data parallelism, its allocation $n$ is supposed to divide its global batch size $B_m$ to avoid resource under-utilization due to uneven partition of samples.
For another example, if an \mop is applied tensor parallelism or sequence parallelism with degree 2, its allocation $n$ is supposed to be divisible by this degree, thus $n=3, 5, 7$ as invalid.
Such \textsl{valid} constraint ensures the allocation plan for each \mop is practical.
Specially, allocation with $\underline{n_m}=0$ is treated as a dummy allocation (e.g., \mop $3$ in Fig.~\ref{fig:mtp_allocator}), which preserves the optimum property of Cond.~\eqref{eq:bp_dscr_alloc_2} but will then be ignored.

Then, we reinstates $l$'s as integers by rounding $\overline{l_m}, \underline{l_m}$ to the nearest integers.
If the rounded $l$ equals $0$, this ASL-tuple will be ignored.
This rounding procedure preserves the integrity of Cond.~\eqref{eq:bp_dscr_alloc_1} and introduces only minor bias to Cond.~\eqref{eq:bp_dscr_alloc_2}.
Finally, the discretized ASL-tuples of all \mops form the allocation plan.
Note that the allocation plan only ensures the longest execution time among all \mops is approximately $\widetilde{C}^\ast$, yet it does not specify the start time for each ASL-tuple, which is determined by wavefront scheduler in \S\ref{subsec:scheduler}.

\subsection{Wavefront Scheduler} \label{subsec:scheduler}

Given the allocation plan from the resource allocator, we now describe how \sname schedules the execution of \mops.
We first introduce the concept of \textsl{\wave}, the scheduling unit of \name.
Then we present our wavefront scheduling, which schedules the execution of \mops greedily for each \wave. 
Finally, the operator dependencies among \mlevels are reinstated by merging the wavefront schedules together.

\subsubsection*{Definition of \textsl{\wave}}
It is worthy to note that, although Theorem~\ref{thm:pcntn} implies that all \mops share the same start and end time in the continuous form, this property does not hold after the discretization process. 
The reason is that the execution time of ASL-tuples may vary, or the resources are insufficient to execute all tuples concurrently.
To cope with this problem, we devise a fine-grained wavefront scheduler that slices the \mops and selects a few of them to execute concurrently on different groups of devices.
We define \textit{\wave} as the smallest scheduling unit, which corresponds to one concurrent execution as aforementioned.
The wavefront scheduler attempts to minimize the device idle time in each \wave, by (1) occupying the devices as many as possible to maximize device utilization (Wavefront Scheduling step {\large \textcircled{\small 1} \textcircled{\small 2}}), and (2) aligning the execution time spans of different sliced \mops to avoid idle time (Wavefront Scheduling step {\large \textcircled{\small 3}}).
As illustrated in Fig.~\ref{fig:mtp_execution_plan}, resource (device) allocation remains unchanged in one \wave, and transmission of data flow occurs only between two \waves.
Next, we introduce our greedy algorithm that crafts the \waves to form the scheduling plan.


\subsubsection*{Wavefront Scheduling}

\begin{algorithm}[!t]
    \caption{\small{Wavefront Scheduling for one \mlevel}}
    \label{alg:scheduling}
    \scalebox{0.98}{
    \begin{minipage}{\textwidth}
    \small
        \LinesNumbered
        \KwIn{
            \# Devices $N$, start time $T_{start}$, \\
            \quad\quad\quad $alloc\_plan=\{m \rightarrow \{\langle \overline{n_m},\cdot,\overline{l_m} \rangle, \langle \underline{n_m}, \cdot, \underline{l_m} \rangle\} \}$ 
        }
        \KwOut{
            Wavefront schedule $P= \bigcup_k \mathcal{S}_k$, end time $T_{end}$
        }
        $T_{current} \gets T_{start}$;
        $P \gets \varnothing$;
        $\setsub{remain} \gets alloc\_plan$\;
        \While(\tcp*[h]{schedule for \wave $k$}){$\setsub{remain}$ is not empty}{
            $\setsub{cand} \gets \texttt{Propose\_Candidate\_Set}(N, \setsub{remain})$\;
            $\setsub{cand} \gets \texttt{Extend\_Resources\_If\_Needed}(\setsub{cand})$\;
            $T_{wave}, \setsub{sched} \gets \texttt{Align\_Time\_Span}(\setsub{cand})$\;
            $\setsub{k} \gets \texttt{Set\_Start\_Time}(\setsub{sched}, T_{current})$;
            $P \gets P \cup \setsub{k}$\;
            $\setsub{remain} \gets \setsub{remain} - \setsub{sched}$;
            $T_{current} \gets T_{current} + T_{wave}$\;
        }
        \KwRet{$P, T_{current}$}
    \end{minipage}
    }
\end{algorithm}

As outlined in Alg.~\ref{alg:scheduling}, the scheduler iteratively crafts \waves in a greedy manner. 
Below we discuss how one \wave is crafted with Fig.~\ref{fig:mtp_execution_plan} as an example.
\begin{enumerate}[label=\large\protect\textcircled{\small\arabic*}, nosep, leftmargin=0pt, labelwidth=-6pt, align=left]
\item First, the scheduler greedily proposes ASL-tuples to form a candidate set, aiming to utilize as many devices as possible (line 3). For instance with Fig.~\ref{fig:mtp_execution_plan}, the scheduler proposes the first ASL-tuple of \mop 1 to craft \wave 1 since it occupies all devices. Similarly, for \wave 2, it proposes the ASL-tuples of \mop 1, 2, and 4, which correspond to 4, 2, 2 devices, respectively, in order to make full use of all devices.
\item If the candidate set fails to occupy all devices, the cluster resources will be underutilized. To address this issue, we extend the allocated resources in specific tuples to ensure all devices are utilized (line 4). 
For instance, in \wave 4 of Fig.~\ref{fig:mtp_execution_plan}, the allocation of \mop 4 is extended from 1 device to 2 devices.
Resource extension is prioritized for \mops with larger remaining execution time, with the hope of balancing the remaining workload among the \mops.
\item In most cases, the proposed ASL-tuples differ in execution time. If we directly craft a \wave with them, it would be inefficient since there must be idle devices. 
Fortunately, this can be avoided by dissecting the ASL-tuples to align their time span (i.e., only a few number of operators in the \mop are scheduled in this \wave). 
For instance, in \wave 2 of Fig.~\ref{fig:mtp_execution_plan}, the proposed ASL-tuples for \mop 1, 2, and 4 correspond to 9, 14, and 3 operators, respectively. To align the execution time, the ASL-tuples for \mop 1 and 2 are dissected, with only 1 and 2 operators of them being scheduled, while the remaining 8 and 13 operators left to be scheduled in subsequent \waves.
Our scheduler simply aligns the time span w.r.t. the ASL-tuple with shortest execution time (e.g., the one for \mop 4 in previous example), and computes the aligned time span as the duration of current \wave (line 5).
\item After the time span alignment, the scheduler concludes the current \wave (lines 6-7), including specifying the start time for operators that are scheduled in this \wave, and removing them from the remaining set.
\end{enumerate}

\subsubsection*{Merging \mlevels}
As stated in \S\ref{subsec:contraction}, \mops are decoupled into \mlevels to disentangle operator dependencies. \name invokes the aforementioned allocation and scheduling for each \mlevel individually, and merges their wavefront schedules together as the final execution schedule.

\subsection{Device Placement} \label{subsec:device_place}

\begin{figure}[!t]
    \centering
    \includegraphics[width=\linewidth]{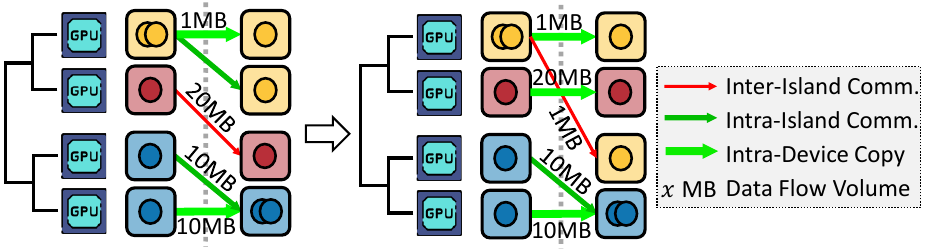}
    \caption{
    Illustration of \name device placement.
    }
    \label{fig:mtp_device_place_c}
\end{figure}

Given the wavefront schedule, which consists of the allocation amount and execution time of each \mop, we now discuss how \name determines the specific devices for each \mop, known as device placement, which affects the inter-\wave communication overhead and memory consumption.
\name employs several guidelines based on empirical insights or observations to optimize device placement.


\subsubsection*{Intra-Device-Island Placement}
Placement within a device island is always preferred for each \mop and data flow between \mops. 
A device island consists of devices connected by high-bandwidth interconnects (e.g., NVLink), typically adjacent devices, such as adjacent GPUs within one node. 
For \mops, prioritizing intra-island placement reduces the potential intra-\mop communication costs. 
For data flow between \mops across \waves, intra-island placement reduces transmission costs leveraging the high intra-island bandwidth or even faster intra-device copying.

\subsubsection*{Prioritizing High Communication Workloads}
When it's infeasible to place all \mops and data flows within the device island, \name will 
estimate the communication volume of each \mop and data flow to prioritize placing those with higher volumes within a device island.
For instance, in Fig.~\ref{fig:mtp_device_place_c}, the data flow volume between red \mops is significantly higher than that between yellow ones. Therefore, \name prefers to place the data flow between red ones within the island.
This guideline ensures the most communication-intensive components receive the most efficient hardware configuration to minimize communication overhead.



\subsubsection*{Device Memory Balance}
As each device holds heterogeneous \mops, the memory overhead varies across devices.
Placing too many memory-intensive \mops on a single device may cause out-of-memory (OOM) errors. 
Therefore, \name actively balance the memory load across all devices during placement.
Specifically, \name estimates each \mop's memory consumption, tracks available memory on devices, and prioritizes placement on the device with the most available memory.
Besides, for \mops sharing the same parameters, we prioritize placing them on the same device to minimize redundant storage.

Based on these guidelines, \name performs device placement \wave by \wave greedily, prioritizing the minimization of communication overhead, such as inter-\wave transmission, while simultaneously maintaining device memory balance. 
When OOM occurs due to imbalanced placement, \name will consider alternative placements with sub-optimal communication costs and better memory balance. If necessary, backtracking is employed to adjust the placements from earlier \waves to effectively address the OOM issues.

\subsection{Runtime Engine} \label{subsec:engine}

\begin{figure}[tbp]
    \centering
    \includegraphics[width=\linewidth]{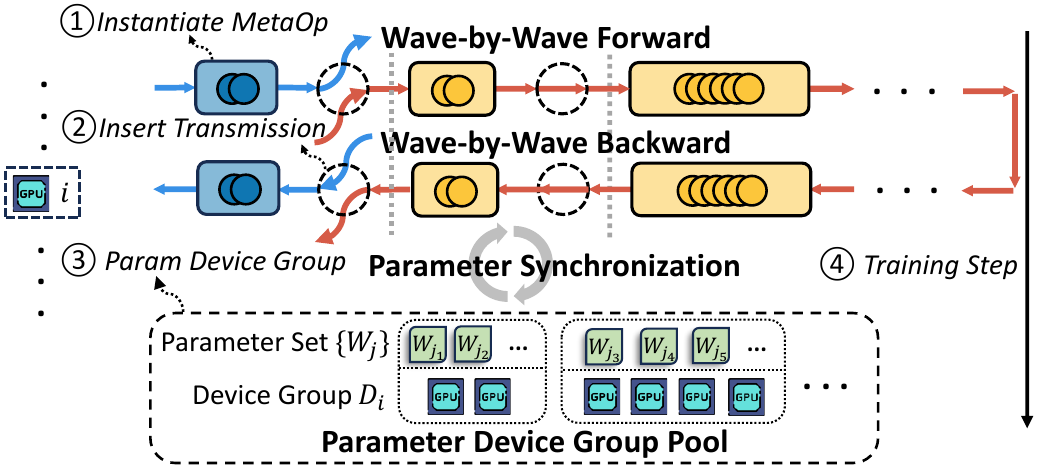}
    \caption{
    Illustration of \name runtime engine.
    }
    \label{fig:mtp_runtime}
\end{figure}

The runtime engine is responsible for running the execution plan to facilitate efficient MT MM training,
which is more complex than conventional single-task training, as each device handles heterogeneous \mops and local computation graphs.
\name runtime engine operates in four main steps:

\begin{enumerate}[label=(\arabic*), nosep, leftmargin=0pt, labelwidth=-6pt, align=left]

\item \textbf{Localization.} 
Initially, \name localizes the execution plan to each device. 
Specifically, each device instantiates the corresponding \mop of each \wave locally, and initializes the required model components and parameters.

\item \textbf{Intra-task Data Dependency.} 
Secondly, \name inserts transmission operators to connect the \mops across \waves to handle the data flow dependencies, 
including forward activations and backward gradients.
According to the devices of \mops and data format requirements, operations such as \textit{copy}, \textit{shard}, \textit{concat}, \textit{send}, and \textit{receive} are used to transmit data flows with minimal overhead.
This step not only correctly handles data flow dependencies between \mops but also links the \mops on each device into a complete local computation graph ready for execution.

\item \textbf{Inter-task Model Dependency.} 
Then, \name manages parameter device groups for synchronization among various tasks by maintaining a global parameter device group pool. 
During each iteration, for each parameter $W_j$, all tasks or modalities that activate it on different devices contribute to its gradient, which needs to be accumulated and synchronized to facilitate parameter sharing. 
Therefore, before the training process, \name scans all devices to determine the device group $D_i$ for each parameter $W_j$, which represents $W_j$ is shared and should be synchronized within group $D_i$.
For efficiency, \name manages parameters with the same device group collectively and maintains a global parameter device group pool $\{D_i\rightarrow \{W_j\}\}$, where each device group $D_i$ corresponds to a set of parameters $\{W_j\}$.

\item \textbf{Training Step.} 
Finally, the training process is ready to begin. In each iteration of \name, each device executes the forward and backward propagation of the local computation graph in a \wave-by-\wave manner, which is comprised of the interleaved execution of \mops and transmission of data flow.
Following the forward and backward phases, \name performs group-wise parameter synchronization to maintain the parameter consistency. 
Specifically, each parameter set $\{W_j\}$ is synchronized within its corresponding device group $D_i$ in the parameter device group pool.

\end{enumerate}

\section{Implementation}


\name is an efficient and scalable MT MM training system built on PyTorch with 10K Loc in Python: 2.1K LoC for the execution planner and 7.9K LoC for the runtime engine.
We implement the data flow transmission with NCCL batched P2P primitives and the parameter device groups with NCCL communication groups.
\name provides the users with simple, user-friendly and flexible API for defining MT MM training workloads.
Specifically, training tasks in \name are represented as \textit{{\sname}Task}, and users can define various multi-modal tasks by customizing PyTorch modules and connecting them flexibly through the \textit{add\_flow} API in \name. 
Alternatively, user can also define different computational logic for various tasks implicitly within a single unified model. 
\name can automatically split the modules and construct \textit{{\sname}Tasks} via PyTorch FX Tracer, streamlining task definition.
After the definition of multi-modal tasks, \name conducts the optimization workflow automatically, as illustrated in Fig.~\ref{fig:mtp_overview}, and the \name runtime engine provides efficient and scalable model training process.

\begin{table}[t]
    \centering
    \caption{Experimental setups.}
    \begin{minipage}{\linewidth}
        \centering
        \subcaption{
           Heterogeneity awareness of system competitors.
        }
        \label{tab:baseline_feature}
        \centering
        \begin{tabular}{c|c|c}
        \toprule
        \textbf{Competitors} & \textbf{Inter-Task} & \textbf{Intra-Task}   \\ \midrule
        Megatron-LM / DeepSpeed & \XSolidBrush    & \XSolidBrush \\ \hline
        \distmm        & \XSolidBrush    & \CheckmarkBold        \\ \hline
        \name-Optimus     & \CheckmarkBold  & \XSolidBrush             \\ \hline
        \name        & \CheckmarkBold  & \CheckmarkBold        \\
        \bottomrule
        \end{tabular}
    \end{minipage}
    
    \vspace{5pt}
    \begin{minipage}{\linewidth}
        \centering
        \subcaption{Configuration of MT MM models for evaluation.}
        \label{tab:model_config}
        \centering
        \begin{tabular}{c|c|c|c}
        \toprule
        \textbf{\makecell{MT MM\\Model}} & \textbf{\makecell{Multitask-\\CLIP}} & \textbf{OFASys} & \textbf{QWen-VAL} \\ \midrule
        \# Param. & 1.20B & 0.66B & 9.25B \\ \hline
        \# Modalities & 6 & 6 & 3 \\ \hline
        \# Tasks & 10 & 7 & 3 \\ \hline
        \makecell{Cross-Modal\\Module} & \makecell{Contrastive\\Loss} & \makecell{Enc-Dec\\LLM} & \makecell{Dec-only\\LLM} \\ \bottomrule
        \end{tabular}
    \end{minipage}
\end{table}

\begin{figure*}[t]
    \centering
    \includegraphics[width=\linewidth]{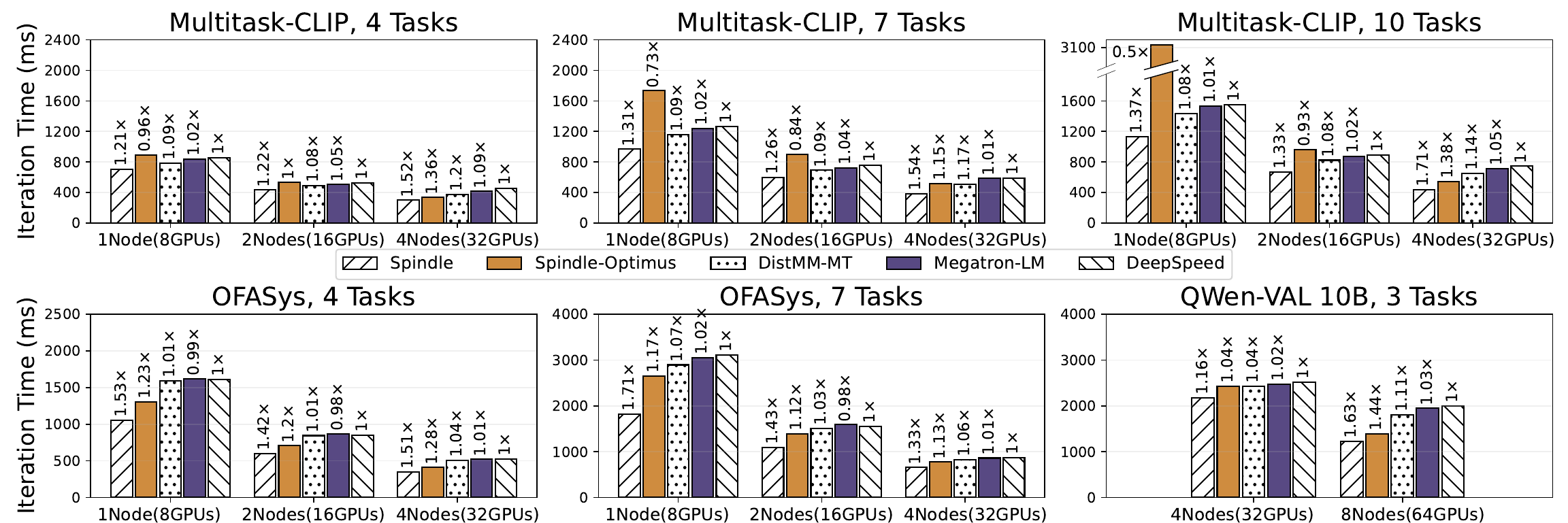}
    \caption{
    End-to-end performance comparison for \name and baseline systems. 
    Shorter bars indicate superior system performance. 
    The numbers above the bars denote each system's speedup compared to DeepSpeed (larger than $1$ is faster).
    }
    \label{fig:end2end-main-result}
\end{figure*}

\section{Experiments}

\subsection{Experimental Setups} \label{subsec:exp_setup}

\subsubsection*{Competitors.}

We compare \name with SOTA (state-of-the-art) distributed training systems, Megatron-LM~\cite{DBLP:megatron} and DeepSpeed~\cite{Deepspeed}. 
We also introduce other two systems that represent typical strategies for multi-task training, considering inter-task and intra-task heterogeneity respectively.
Tab.~\ref{tab:baseline_feature} summarizes the features of competitors.

\noindent (1)\&(2) \textbf{Megatron-LM \& DeepSpeed:}
Megatron-LM~\cite{DBLP:megatron} and DeepSpeed~\cite{Deepspeed} are widely used SOTA training systems tailored for single-task training. 
The na{\"i}ve approach to train MT MM models on these systems is to decouple all sub-models on separate devices (\S\ref{sec:intro}), which requires plenty of resources and is impractical. 
Therefore, we decouple sub-models on temporal dimension within each iteration, where each sub-model takes up the whole cluster within a short time period, and is dependently and sequentially executed.




\noindent (3) \textbf{\distmm:}
DistMM~\cite{distmm-nsdi} is a recent training system designed for multi-modal models, but focusing on single task only.
\distmm represents a multi-task (MT) extension of DistMM.
It decouples multi-tasks, and for each single MM task allocates appropriate resources to different multi-tower modality encoders.
Then it executes tasks sequentially.


\noindent (4) \textbf{\name-Optimus:}
This baseline represents a workload-aware task-level resource allocation strategy, which adapts allocations according to the workload at the task level granularity.
It's inspired by Optimus~\cite{DBLP:Optimus}, an effective cluster job scheduling system which proposes a greedy resource allocation scheme and iteratively assigns devices to the job that has the largest marginal gain.
Despite differences between job scheduling and multi-task training (\S\ref{sec:related_work}), we apply a similar principle and devise the marginal gain as $(T^\comp_m(n) - T^\comp_m(n')/(n'-n)$, i.e., the task completion time reduction scaled by the allocation increment from $n$ to $n'$.
Here $n'$ is the next valid allocation number larger than $n$.

\subsubsection*{Experimental Workloads}

As illustrated in Tab.~\ref{tab:model_config}, we select three models to represent popular MT MM workloads and conduct experiments on these workloads.

\noindent (1)
\textbf{Multitask-CLIP}, which adopts the same structure of ImageBind~\cite{DBLP:ImageBind}, is a multi-task variation to the classic and pioneer CLIP~\cite{DBLP:CLIP} model.
Many multi-modal models~\cite{DBLP:CLIP,DBLP:ALIGN,DBLP:Florence,DBLP:VideoCLIP,DBLP:Audioclip,DBLP:ImageBind} follow this paradigm for multi-task training.
Its cross-modal module (contrastive loss), has much smaller workload compared to its modality encoder, where most computation occurs.

\noindent (2)
\textbf{OFASys}~\cite{DBLP:OFASys} further generalizes the MT MM paradigm, using a unified LM of encoder-decoder structure for cross-modal processing. In OFASys, the cross-modal module's workload is comparable to that of the modality encoders.

\noindent (3)
\textbf{QWen-VAL}~\cite{DBLP:Qwen-VL,DBLP:Qwen-Audio} adopts a modern, compute-intensive decoder-only LLM, with the workload of the cross-modal module usually larger than modality encoders.
Recent multi-modal models like SPHINX-X~\cite{liu2024sphinx-x}, DeepSpeed-VisualChat~\cite{yao2023deepspeed-visualchat}, and BLIP-2~\cite{DBLP:BLIP-2}, employ this structure.

These workloads effectively represent the majority of MT MM workloads (and different workload distribution between modality encoders and cross-modal modules in Fig.~\ref{fig:intro_observation}), regardless of specific model structure variations.
\subsubsection*{Protocols}
We conduct experiments on an 8-node GPU cluster. Each node consists of 8 NVIDIA A800 80 GB GPUs equipped with NVLink, and the nodes are interconnected by 400 Gbps InfiniBand. 
Since the baseline systems do not support automatic planning given a targeted MT MM model training workload, we manually tune their parallel configurations and memory optimization techniques (e.g., data and tensor parallelism degree, ZeRO stage, activation checkpointing, and etc.) to achieve the best performance.
Averaged training time over 100 iterations is reported.

\subsection{End-to-End Performance}

Fig.~\ref{fig:end2end-main-result} displays end-to-end iteration time comparisons between \name and baseline systems across various model workloads, multi-modal task configurations, and cluster sizes.

\subsubsection*{Comparison with SOTA systems.}
In general, compared to SOTA training systems, i.e., Megatron-LM and DeepSpeed, \name achieves speedup ratios of up to 67\% and 71\%, respectively.
Below we delve into details. 

To begin with, \name consistently outperforms the competitors across different task configurations. Notably, \name excels when handling a larger number of tasks.
On the 10-task Multitask-CLIP and 7-task OFASys workloads, \name achieves speedup ratios ranging from 31\% to 71\% compared to SOTA systems.
This underscores \name's excellent scalability with increasing task counts.

In addition, \name consistently achieves optimal performance across various cluster sizes.
On Multitask-CLIP, \name achieves the highest speedup ratios of 37\%, 33\%, and 71\% on 8, 16, and 32 GPUs, respectively.
Notably, \name maintains high efficiency even when the scalability of SOTA systems begins to diminish --- that is, when the increase in resources does not correspond to significant speed improvements.
For example, in 4-task Multitask-CLIP, expanding the cluster size from 16 to 32 GPUs results in only modest speedup of up to 1.21$\times$ for SOTA systems, whereas \name still achieves a 1.45$\times$ speedup.
This efficiency stems from \name's heterogeneity-aware and operator-level fine-grained resource allocation and scheduling.
For example, for a lightweight audio operator, DeepSpeed needs to parallelize it on the whole cluster with 16 GPUs due to its workload-unaware nature, causing the computational kernel to be underutilized or even idle, while \name may parallelize it with only 4 GPUs to ensure high utilization.
Besides, when scaling from 16 to 32 GPUs, \name may maintain 4-GPU-allocation for the lightweight operator to keep high utilization.


More importantly, \name also exhibits excellent scalability w.r.t. model size. 
On larger model QWen-VAL, \name achieves a maximum speedup of 1.16$\times$ on 32 GPUs and 1.63$\times$ on 64 GPUs.
Notably, when training QWen-VAL, \name achieves a 1.78$\times$ speedup when scaling from 32 to 64 GPUs, whereas SOTA systems only achieve up to 1.27$\times$ speedup.
This is unsurprising since \name allows flexible allocations to avoid the unsatisfactory scalability of \mops with light workloads, as discussed above and in \S\ref{subsec:scale_est}.



\begin{figure}[!t]
    \centering
    
    \begin{minipage}{\linewidth}
        \centering
        \includegraphics[width=\linewidth]{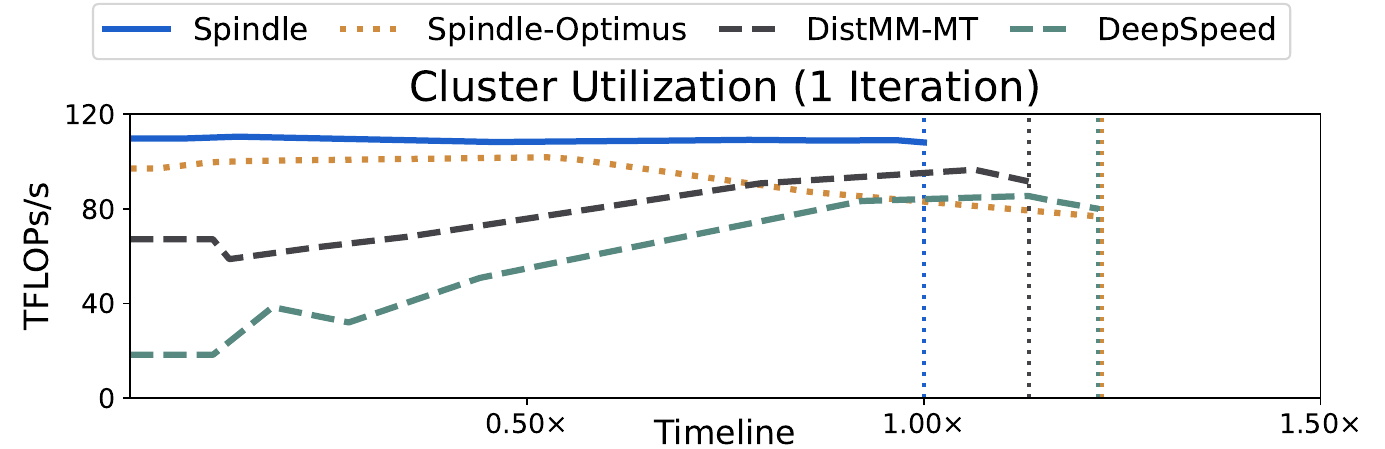}
        \subcaption{Average cluster utilization over time within one iteration.
        Higher positions on the y-axis indicate higher utilization.
        }
        \label{fig:case_study_timeline}
    \end{minipage}
    \begin{minipage}{\linewidth}
        \centering
        \includegraphics[width=\linewidth]{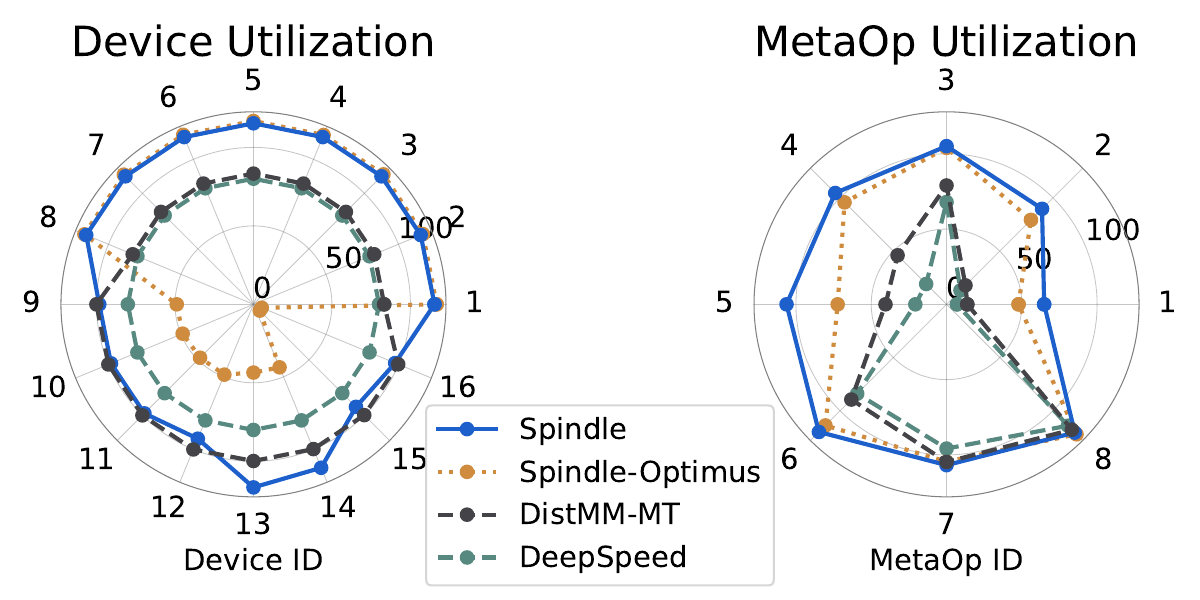}
        \subcaption{Utilization of each device and each \mop.
        Points closer to the outer edge of the spider chart represent higher utilization.
        }
        \label{fig:case_study_metaop_device}
    \end{minipage}
    \caption{
    Case study of Multitask-CLIP (4 tasks, 16 GPUs). 
    Utilization is measured in computation FLOPs per second.
    }
    \label{fig:case_study}
\end{figure}

\subsubsection*{Comparison with other baselines}
Next, we discuss the performance of other baselines, i.e., \name-Optimus with task-level resource allocation strategies, and \distmm with the single-task strategy for MM models.

\name-Optimus, employing workload-aware task-level resource allocation, exhibits great performance, especially in larger-scale clusters, with the speedup ratio up to 44\% compared to DeepSpeed.
However, there are still many scenarios where it underperforms.
Its task-level strategy overlooks the workload heterogeneity within tasks, thereby limiting training efficiency. 
Moreover, the coarse granularity at task level can sometimes fail to achieve ideal load balancing among tasks, causing many devices to become idle during the latter part of the iteration. 
In comparison, \name enables finer-grained strategy, with operator-level resource allocation and wavefront load balancing, consistently achieving higher efficiency compared to \name-Optimus.

\distmm also performs better than SOTA systems in most cases, with the speedup up to 20\%, benefiting from its intra-task workload awareness and resource allocation.
However, it's designed for single-task MM models, which decouples tasks and optimizes each one separately, making it far from achieving the global optimum in multi-task cases.
For OFASys, \distmm shows poor performance.
This is because \distmm gains acceleration by parallelizing sub-models in the same task.
However, OFASys uses a lightweight text adaptor, so most tasks that pair a modality with text are dominated by the other modality, making the sub-model parallelization ineffective.
Compared to \distmm, \name jointly optimizes the allocation and scheduling of all tasks and operators, therefore consistently outperforming the single-task strategy of \distmm, achieving a speedup ratio of up to 59\%.





\subsection{Case Study}

To better understand the advantages and performance gain of \name over the other competitors, we further conduct an in-depth case study of Multitask-CLIP (4 tasks, 16 GPUs). Fig.~\ref{fig:case_study} presents system performance considering three key metrics: cluster average utilization over time, average utilization per device, and computational utilization of each \mop.

Firstly, DeepSpeed, representing SOTA systems, which executes the tasks sequentially with all resources, experiences fluctuating utilization due to the workload heterogeneity, leading to generally low overall utilization. 
\name-Optimus, which allocates resources at task level, improves cluster utilization at the iteration beginning, but as tasks with light workloads finish, more devices become idle, declining overall utilization. 
\distmm manages to enhance utilization via intra-task resource allocation, but the ignorance of inter-task heterogeneity limits its utilization.
In contrast, \name maintains consistently high utilization throughout the iteration. 


Furthermore, \name significantly elevates the utilization of all devices and \mops, showcasing its superior handling of workload balance via operator-level strategies. 
In contrast, DeepSpeed shows lower utilization across all devices and \mops. 
Although task-level strategies of \name-Optimus can enhance the utilization of certain devices, the coarse granularity of allocation inevitably leads to workload imbalances, leaving many devices underutilized.
\distmm also improves the utilization of certain devices and \mops, but the results are still unsatisfactory as it fails to reach the global optimal parallel plan for multi-tasks.
Overall, \name's unified optimization captures a close-to-optimal execution plan with workload balance, leading to consistently high utilization in all aspects.

\begin{figure}[tbp]
    \centering%
    \includegraphics[width=\linewidth]{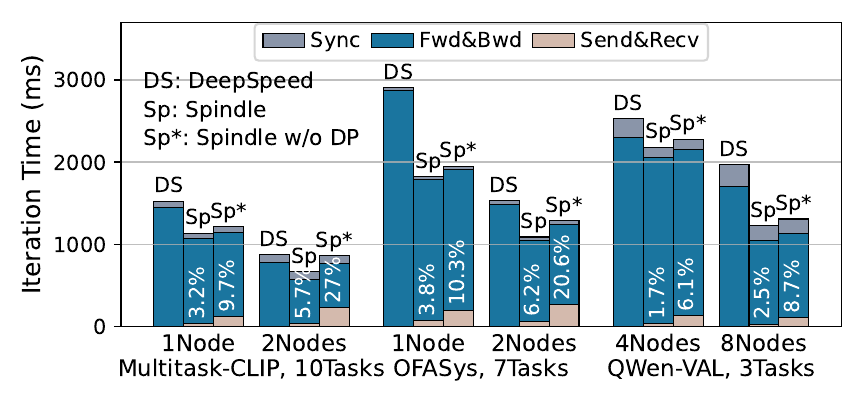}
    \caption{
        Time breakdown analysis.
        Percentage of inter-\wave \textit{send} \& \textit{receive} in total time is labeled for ablation study.
    }
    \label{fig:time_breakdown}
\end{figure}

\subsection{Time Breakdown}

Fig.~\ref{fig:time_breakdown} shows the runtime breakdown for \name and DeepSpeed across various workloads, primarily consisting of forward and backward propagation, parameter synchronization, and inter-\wave \textit{send} and \textit{receive}. We've isolated parameter synchronization from the backward phase for individual analysis.
In MT MM training, the forward and backward propagation dominates the runtime, typically accounting for 80\%-95\% due to the large number of tasks and computational demands. \name focuses on reducing this significant time component through flexible resource allocation and scheduling. 
Parameter synchronization usually consumes a small fraction of the time, about 5\%-15\%, since it only occurs after accumulating gradients from multiple tasks.
Furthermore, while \name introduces extra overhead for inter-\wave \textit{send} and \textit{receive}, this overhead remains minimal, typically not exceeding 6\%, thanks to the device placement mechanism that avoids unnecessary communications. 

\begin{figure}[tbp]
    \centering
    \centering
    \includegraphics[width=\linewidth]{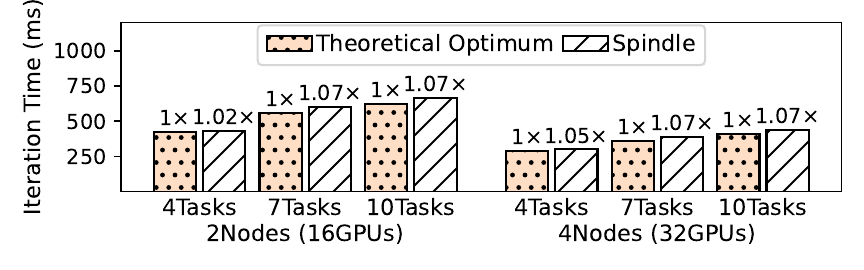}
    \caption{
        Optimality analysis of \name execution planner. 
        Evaluated on Multitask-CLIP. 
    }
    \label{fig:optimality}
\end{figure}

\subsubsection*{Ablation on Device Placement} \label{subsubsec:ablation_device_place}
We conduct an ablation study on the device placement strategy in \S\ref{subsec:device_place}, focusing on its impact on inter-\wave communication overhead, which is the extra overhead of our system. 
Specifically, we compare it with a sequential placement strategy, which assigns each \mop with consecutive devices. 
In Fig.~\ref{fig:time_breakdown}, our results indicate that the inter-\wave communication overhead of the sequential placement strategy is approximately 3-6 times greater than that of \name, taking up to 27\% of the end-to-end training time.
However, \name's placement strategies reduces this overhead to 6\%.
This demonstrates the effectiveness of our locality-aware placement, which significantly reduces the extra communication overhead.

\subsection{Execution Planner Evaluation}
\subsubsection*{Optimality Analysis}
We analyze the optimality of \name execution planner in Fig.~\ref{fig:optimality} by comparing the iteration time to the theoretical optimum $\widetilde{C}^\ast$ derived from Theorem~\ref{thm:pcntn}. 
Although unachievable due to the relaxed constraints (\S\ref{subsec:allocator}), $\widetilde{C}^\ast$ serves as a theoretical performance upper bound. 
The \name execution planner preserves most of the optimum property when finding the practical solution (e.g., Cond.~\eqref{eq:bp_dscr_alloc_1} ~\eqref{eq:bp_dscr_alloc_2}), but still introducing minor biases (e.g., reinstating $l's$ to integers in \S\ref{subsec:allocator}, resource extension in \S\ref{subsec:scheduler}). 
In Fig.~\ref{fig:optimality}, we find that across various configurations, the deviation between \name and theoretical optimum is consistently low, below 7\%. 
This observation underscores the effectiveness of \name in offering a practical and near-optimal execution plan for MT MM models.

\begin{figure}[t]
    \centering
    \includegraphics[width=\linewidth]{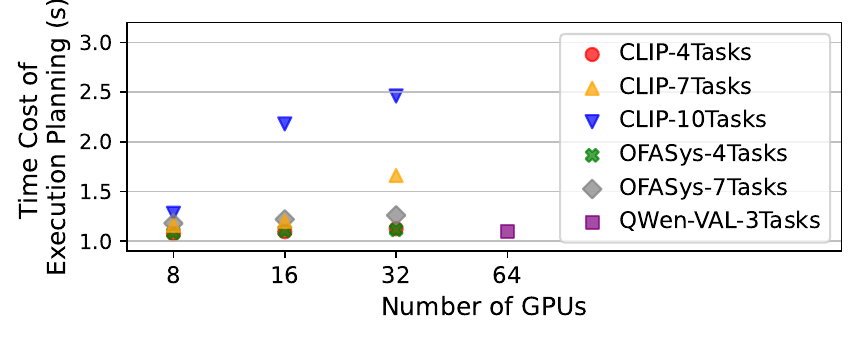}
    \caption{
    Time cost (s) of \name's execution planner.
    }
    \label{fig:complexity}
\end{figure}

\subsubsection*{Complexity Analysis}
We briefly analyze the complexity of execution planner.
Given \mops' scalability curves, the planning process consists of three parts: resource allocation, wavefront scheduling, and device placement.
Among the first two parts, most of the time is spent on solving the continuous optimization problem via bisection search \ifasplos\else{(Appendix~\ref{sec:append_p_cntn}). }\fi
In comparison, the complexity of wavefront scheduling is relatively small, scaling linearly with the number of waves, which is at most twice the number of \mops.
This is because each wave consumes all layers of at least one ASL-tuple (\S\ref{subsec:scheduler} {\large \textcircled{\small 3}}), while each \mop produces two ASL-tuples (\S\ref{subsec:allocator} bi-point discretized allocation).
The third part, device placement, uses a constrained-depth recursive search with simple heuristics.
The searching time may vary, but is generally within an acceptable range.
As shown in Fig.~\ref{fig:complexity}, \name effectively generates the execution plans within 3 seconds across all experiments.
Moreover, the plan will only be regenerated when input data workload changes, which is not very often compared to the overall training process.
\subsubsection*{More Experimental Results.}
Due to the space constraint, we put more experimental results and analysis in the \ifasplos\else{Appendix}\fi, including more details of experimental workloads in Appendix~\ifasplos\else{\ref{sec:append_workloads}}\fi, evaluation of dynamicity performance in Appendix~\ifasplos\else{\ref{sec:append_dynamic}}\fi, larger-scale simulations in Appendix~\ifasplos\else{\ref{sec:append_largescale}}\fi, comparison on single-task multi-modal (STMM) workloads in  Appendix~\ifasplos\else{\ref{sec:append_stmm}}\fi, memory consumption analysis in Appendix~\ifasplos\else{\ref{sec:append_memory}}\fi, and system implementation performance evaluation in Appendix~\ifasplos\else{\ref{sec:append_sysimple}}\fi.
Please kindly refer to our Appendix for more details.

\section{Related Works} \label{sec:related_work}
\subsubsection*{Cluster Scheduling for DL Jobs}
GPU clusters often design cluster schedulers to coordinate resource allocation and the execution order among multiple DL jobs. 
Some cluster schedulers~\cite{DBLP:Tiresias,DBLP:Gandiva} allocate resources to jobs based directly on user-specified requirements. 
Others~\cite{DBLP:Optimus,DBLP:SLAQ,DBLP:Gavel,DBLP:AntMan,DBLP:Themis,DBLP:Pollux,lyra-eurosys23,heet-asplos24} automatically allocate resources to each job based on the job scalability to the computing resource, many of them minimizing job completion time (JCT). 
For instance, Optimus~\cite{DBLP:Optimus} introduces the marginal gain to guide resource allocation, aiming to minimize job completion time.
Here, we highlight the difference of these works and MT MM model training. 
Unlike the independence among jobs in cluster scheduling, MT MM training involves execution dependencies among tasks.
Furthermore, while traditional scheduling focuses on job-level allocation, MT MM training requires finer-grained strategies to address intra-task workload heterogeneity.



\subsubsection*{Training Optimization on Heterogeneous Cluster}
This line of research focuses on optimizing the distributed training efficiency of DL models on heterogeneous GPU clusters~\cite{DBLP:HetPipe,DBLP:AccPar,DBLP:SDPipe,DBLP:AMP,DBLP:conf/sigmod/MiaoNSYJM021,whale-atc22}. 
While these works primarily concentrate on optimizing single model training and address hardware heterogeneity, \name mainly focus on more complex MT MM models, and addresses the challenges posed by the workload heterogeneity of MT MM models.

\section{Conclusion}

Efficient training of MT MM models faces significant system challenges due to the workload heterogeneity and complex execution dependency. 
In this paper, we proposed \sname to facilitate efficient training of MT MM models via wavefront scheduling, which jointly optimizes heterogeneity-aware workload parallelization and dependency-driven execution scheduling.
Extensive experiments demonstrate the consistent superior performance of \sname, outperforming existing SOTA training systems with speedup ratio up to 71\%.

\begin{acks}
This work is supported by National Science and Technology Major Project (2022ZD0116315), National Natural Science Foundation of China (U23B2048, 62402011), Beijing Municipal Science and Technology Project (Z231100010323002), China National Postdoctoral Program for Innovative Talents (BX20230012), China Postdoctoral Science Foundation (2024M750103), Beijing Natural Science Foundation (4244080), research grant No. IPT-2024JK29, Alibaba-PKU joint program, and High-performance Computing Platform of Peking University. Fangcheng Fu, Xupeng Miao and Bin Cui are the corresponding authors.
\end{acks}

\bibliographystyle{ACM-Reference-Format}
\balance
\bibliography{my-reference}

\ifasplos
\else
\appendix
\clearpage

\section{Details of Scalability Estimator} \label{sec:append_estimator}
\name characterizes the execution time of \mop $m$ over $n$ devices, $T_m(n)$, by a generalized piecewise $\alpha\myhyphen\beta$ function:
{
\small
\begin{equation*} 
T_m(n) = \alpha_{m,i} + {\beta_{m,i} \times {c_m}} + {{\beta_{m,i}'} \times {w_m}}/{n},\forall n \in [n_{i-1}, {n_i}], i = 1,\dots,k
\end{equation*}
}
where $k$ is the number of pieces, $\alpha_{m,i}$ represents the coefficient of fixed overheads (e.g., kernel launch costs), $\beta_{m,i}$ and $\beta_{m,i}'$ represent the reciprocal of execution efficiency (e.g., GPU computation speed and network bandwidth), ${w_m/n}$ denotes the distributed workload of \mop $m$ across $n$ devices (e.g., computational workload), and ${c_m}$ denotes the workload that doesn't scale with $n$ (e.g., communication volume of data parallelism).
Such piecewise function indicates that under varying resource scales, due to changes in the per-device workload, coefficients such as 
$\alpha$, $\beta$ and $\beta'$ might differ, as the invoked kernels may vary across different workloads.

\section{Details of Bisection Search for Optimum of Continuous Problem}\label{sec:append_p_cntn}
Alg.~\ref{alg:bisection_search} illustrates our bisection search algorithm to solve the optimum of malleable project scheduling problem, MPSP.
The function \texttt{Find\_Inverse\_Value($T_m, \widetilde{C}=\frac{\widetilde{C}_{mid}}{L_m}$)} finds the value of $T_m^\inv (\widetilde{C})$.
It first finds the closest valid allocations of \mop $m$, denoted as $\underline{n_m}$ and $\overline{n_m}$, such that $ \widetilde{C} \in [T_m(\underline{n_m}),T_m(\overline{n_m})]$.
It then returns
\begin{equation}
\small
    n_m = \frac{(\widetilde{C} - T_m(\underline{n_m})) \cdot \overline{n_m} +  (T_m(\overline{n_m}) - \widetilde{C}) \cdot \underline{n_m}}{T_m(\overline{n_m}) - T_m(\underline{n_m})},
\end{equation}
which is the linear combination of $\underline{n_m}, \overline{n_m}$
such that $T_m(n_m) = \widetilde{C}$.

\begin{algorithm}[tbh]
    \caption{Bisection Search for MPSP}
    \label{alg:bisection_search}
    \begin{minipage}{\textwidth}
        \LinesNumbered
        \KwIn{
            \# Devices $N$, \\
            \quad\quad\quad linear-piecewise execution time functions $\{T_m\}_{m=1}^M$
        }
        \KwOut{Optimum $P_{MPSP}=\{m \rightarrow \langle n_m^\ast, 0, L_m \rangle \}_{m=1}^M$}
        $\mathcal{T}_{min}, \mathcal{T}_{max} \gets \{ T_m(N) \cdot L_m \}_{m=1}^M, \{ T_m(1) \cdot L_m \}$\;
        $\widetilde{C}_{low}, \widetilde{C}_{high} \gets \operatorname{max} \mathcal{T}_{min}, \operatorname{sum} \mathcal{T}_{max}$\;
        \While{$\widetilde{C}_{high} - \widetilde{C}_{low} > \varepsilon$}{
            $\widetilde{C}_{mid} \gets (\widetilde{C}_{low}+\widetilde{C}_{high})/2$\;
            $P_{MPSP} \gets \{ m \rightarrow \texttt{Find\_Inverse\_Value}(T_m, \frac{\widetilde{C}_{mid}}{L_m})) \}_{m=1}^M$\;
            \eIf{sum of allocations in $P_{MPSP} < N$}{
                $\widetilde{C}_{high} \gets \widetilde{C}_{mid}$\;
            }{
                $\widetilde{C}_{low} \gets \widetilde{C}_{mid}$\;
            }
        }
        \KwRet{$P_{MPSP}$}
    \end{minipage}
\end{algorithm}


\section{Details of Experimental Workloads} \label{sec:append_workloads}
Here, we introduce our experimental workloads in detail.
\noindent (1) \textbf{Multitask-CLIP:}
Multitask-CLIP is a generalized version of CLIP~\cite{DBLP:CLIP}, which extends CLIP to 6 modalities and multiple contrastive learning tasks of paired data modalities.
We utilize the same model structure and configuration of ImageBind~\cite{DBLP:ImageBind}.
We select 6 modalities as ImageBind, including text, vision, audio, motion, thermal, and depth, and select 10 different multi-modal contrastive learning tasks for evaluation, each with distinct workloads.
Each task activates two modality encoders simultaneously without data flow dependency.

\noindent (2) \textbf{OFASys:}
OFASys~\cite{DBLP:OFASys} is a more general MT MM training workload, allowing modalities and tasks to activate the model components flexibly as needed. 
OFASys supports multiple modalities, such as text, vision, audio, motion, box, structure, and supports various multi-modal tasks, such as text summarization, image captioning, visual grounding, speech recognition, text-to-SQL and etc.
OFASys utilizes modality-specific adaptors for different modalities, e.g., ViT for vision data, and adopts a unified encoder-decoder LM with generative loss.
We select 7 different multi-modal tasks for evaluation.

\noindent (3) \textbf{QWen-VAL:}
QWen-VAL is a larger-scale MT MM model with up to 9.25 billion parameters, supporting three modalities, including text, vision, and audio. 
It adopts the same structure and configuration of the popular open-sourced multi-modal LLMs, QWen-VL~\cite{DBLP:Qwen-VL} and QWen-Audio~\cite{DBLP:Qwen-Audio}. 
It has modality encoders for audio and vision, and the extracted modality-specific features are combined with text tokens and together fed into the unified LLM, QWen~\cite{DBLP:QWen}.
We select three tasks for evaluation, i.e., vision-language (VL) task, audio-language (AL) task, and vision-audio-language (VAL) task, representing different combinations of modalities.

\section{Dynamicity Performance}\label{sec:append_dynamic}

\begin{figure}[ht]
    \centering
    \includegraphics[width=\linewidth]{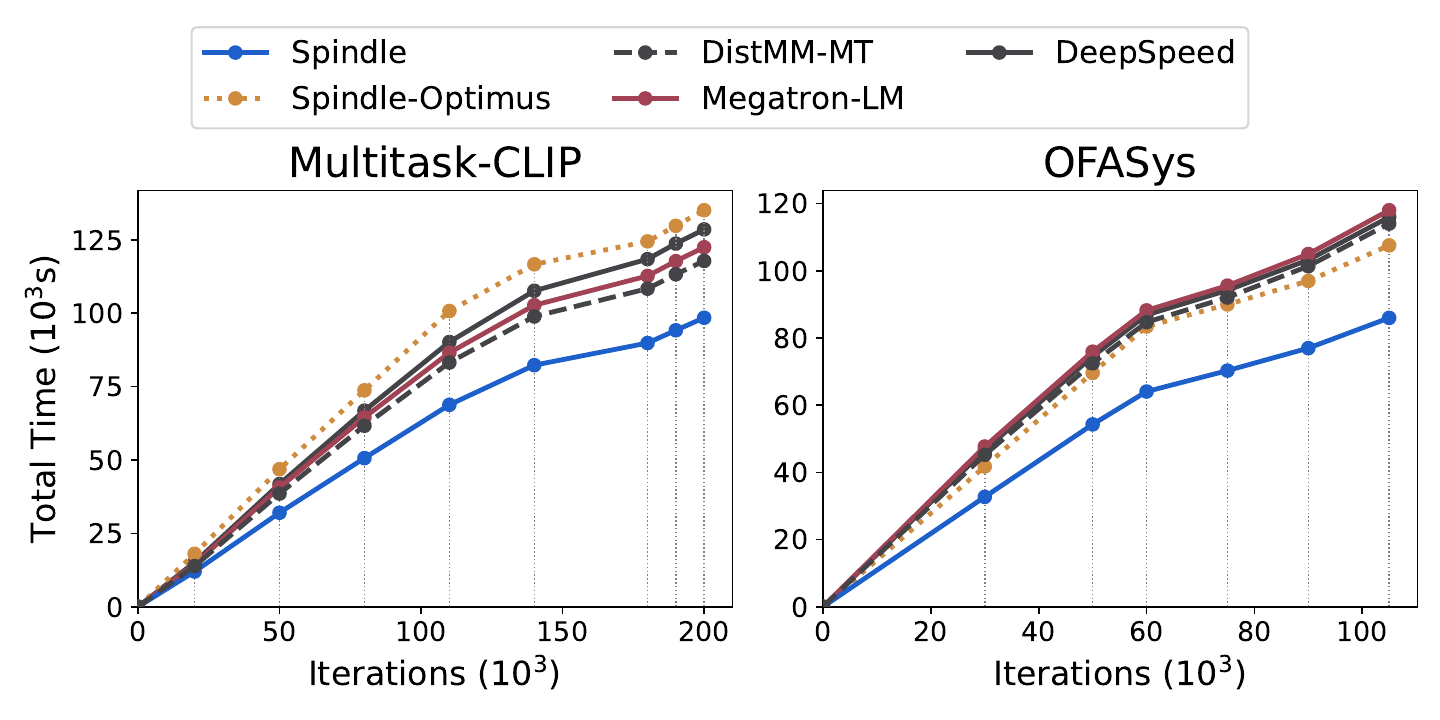}
    \caption{
    Comparison on dynamic multi-task workloads.
    Dots on the curve mark the points when the multi-task workload changes.
    }
    \label{fig:dynamic_workloads}
\end{figure}

We evaluate the performance of various systems during dynamic changes of the multi-task workloads, a common occurrence in MT MM training. For instance, tasks with fewer training data may exit early, and new tasks may join partway through training. We simulate these dynamic changes by altering the training task set.
When the multi-task workloads change, the current model is first saved, and the new set of tasks and the saved model is loaded to continue training.
Fig.~\ref{fig:dynamic_workloads} illustrates the performance of each system under such conditions. \name consistently achieves optimal training efficiency and the shortest overall training time. This advantage is due to \name's adaptability to dynamically changing workloads, enabling it to adopt an appropriate execution plan for the efficient training of MT MM models.

\section{Larger-scale Simulations} \label{sec:append_largescale}
We further conduct evaluation on larger-scale QWen-VAL models (i.e., 30B, 70B).
Due to the lack of massive scale resources, we use the simulation-based approach to estimate the system performance on larger cluster with 256 GPUs.
The speedup ratio of each system compared to DeepSpeed is summarized in Tab.~\ref{tab:simulate-large-scale}.
As can be seen, on larger-scale models and clusters, \name consistently achieves a substantial speedup of over 1.3$\times$ compared to DeepSpeed, validating its scalability w.r.t model size.
In contrast, other competiters achieve speedups of less than 1.1$\times$ compared to DeepSpeed.
Besides, we also want to clarify that MT MM models within 10B are popular and widely deployed in real-world applications (e.g., BLIP-2~\cite{DBLP:BLIP-2}, MiniGPT-4~\cite{DBLP:MiniGPT-4}, QWen-Audio~\cite{DBLP:Qwen-Audio}, DeepSeek-VL~\cite{lu2024deepseekvl}).
Their popularity is not only because many of them are open-source, but also because the required hardware is more accessible in the real world.

\begin{table}[h]
    \centering
    \caption{
    Simulated iteration time speedup over DeepSpeed on larger-scale QWen-VAL workloads (3 Tasks) and larger-scale clusters (256 GPUs).
    }
    \label{tab:simulate-large-scale}
    \begin{tabular}{c|c|c}
    \toprule
    \textbf{Systems} & \textbf{QWen-VAL 30B}  & \textbf{QWen-VAL 70B}  \\ \midrule
    \name           & \textbf{1.34$\times$}  & \textbf{1.36$\times$}  \\ \hline
    \name-Optimus   & 1.05$\times$  & 1.09$\times$  \\ \hline
    \distmm         & 1.04$\times$  & 1.04$\times$  \\ \hline
    DeepSpeed       & 1$\times$     & 1$\times$     \\
    \bottomrule
    \end{tabular}
\end{table}

\section{Comparison on Single-Task Multi-Modal Workload}\label{sec:append_stmm}

We also compare \name with baseline systems on single-task (ST) multi-modal (MT) scenario, which is a special case of MT MM training, as shown in Fig.~\ref{fig:end2end-clip-1task}. We are pleased to observe that even in the single-task scenario, \name outperformed SOTA systems by up to 48\%. This is attributed to \name's fine-grained, operator-level resource allocation and scheduling, which recognize not only the inter-task workload heterogeneity but also intra-task operator workload variations --- a capability beyond the reach of task-level strategies as well as SOTA systems.
It's worth noting that \distmm has similar performance to \name on ST MM scenario, which is reasonable, as \distmm is specifically designed for single-task multi-modal workloads.

\begin{figure}[h]
    \centering
    \vspace{-5pt}
    \includegraphics[width=\linewidth]{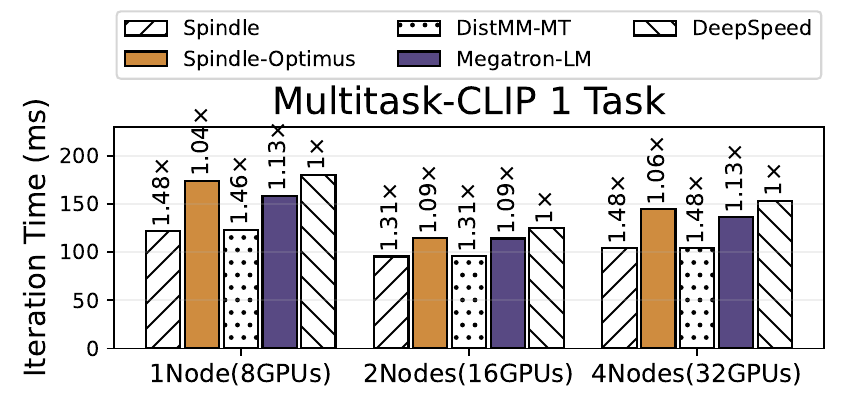}
    \vspace{-15pt}
    \caption{
    End-to-end performance comparison for \sname and baseline systems on 1-task Multitask-CLIP workload.}
    \vspace{-5pt}
    \label{fig:end2end-clip-1task}
    
\end{figure}

\section{Memory Consumption}\label{sec:append_memory}

We also conduct a comparative analysis of memory consumption between \name and the other competitors. Fig.~\ref{fig:case_study_memory} depicts the memory usage for each device in the scenario of Multitask-CLIP (4 tasks, 16 GPUs). Our findings indicate that \name generally exhibits lower memory consumption than SOTA systems such as Megatron-LM and DeepSpeed. This efficiency stems from \name's operator-level strategy and selective parameter storage feature, where only devices that activate a specific operator need to maintain its corresponding parameters, thereby minimizing redundant storage. Additionally, we've observed that task-level strategy \name-Optimus experiences significant memory imbalances. 
In contrast, \name maintains an excellent balance of memory consumption across devices, a success that is attributed to our device placement strategies.

\begin{figure}[h]
    \centering
    \includegraphics[width=.75\linewidth]{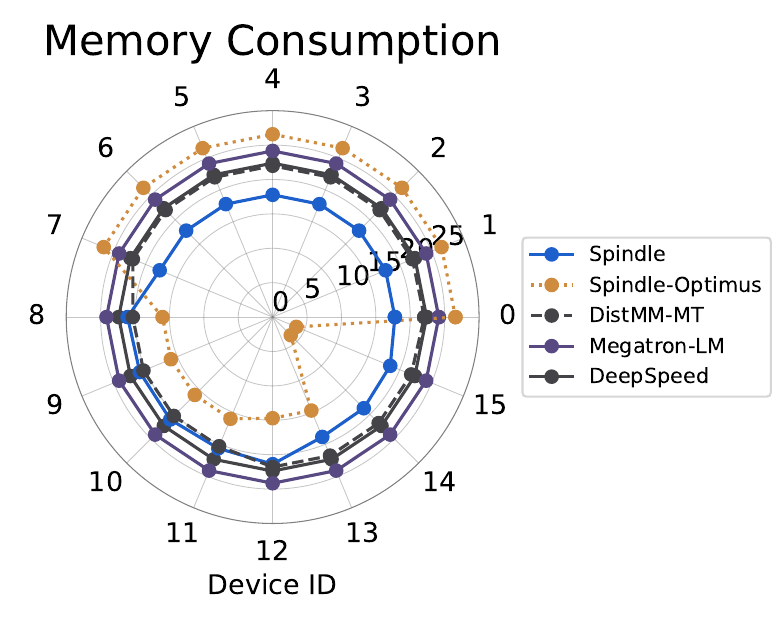}
    \caption{
    Memory consumption (GB) of each device in Multitask-CLIP (4 tasks, 16 GPUs).
    Points closer to the inner edge of the spider chart represent lower GPU peak memory usage.
    }
    \label{fig:case_study_memory}
\end{figure}

\section{System Implementation Performance} \label{sec:append_sysimple}
We validate the system implementation performance of \name.
Specifically, we implement a simple decoupled baseline on \name, \name-Seq, which allocates available devices to each task and execute tasks sequentially within each iteration, similar to SOTA systems, Megatron-LM and DeepSpeed. 
It reflects the implementation performance of our system without specific optimizations for MT MM workloads, i.e., without \name's flexible resource allocation and scheduling strategies.
As shown in Fig.~\ref{fig:seq_comparison}, we find that \name-Seq has similar system performance compared to SOTA systems without specific optimization for MT MM workloads.

\begin{figure}[h]
    \centering
    \includegraphics[width=\linewidth]{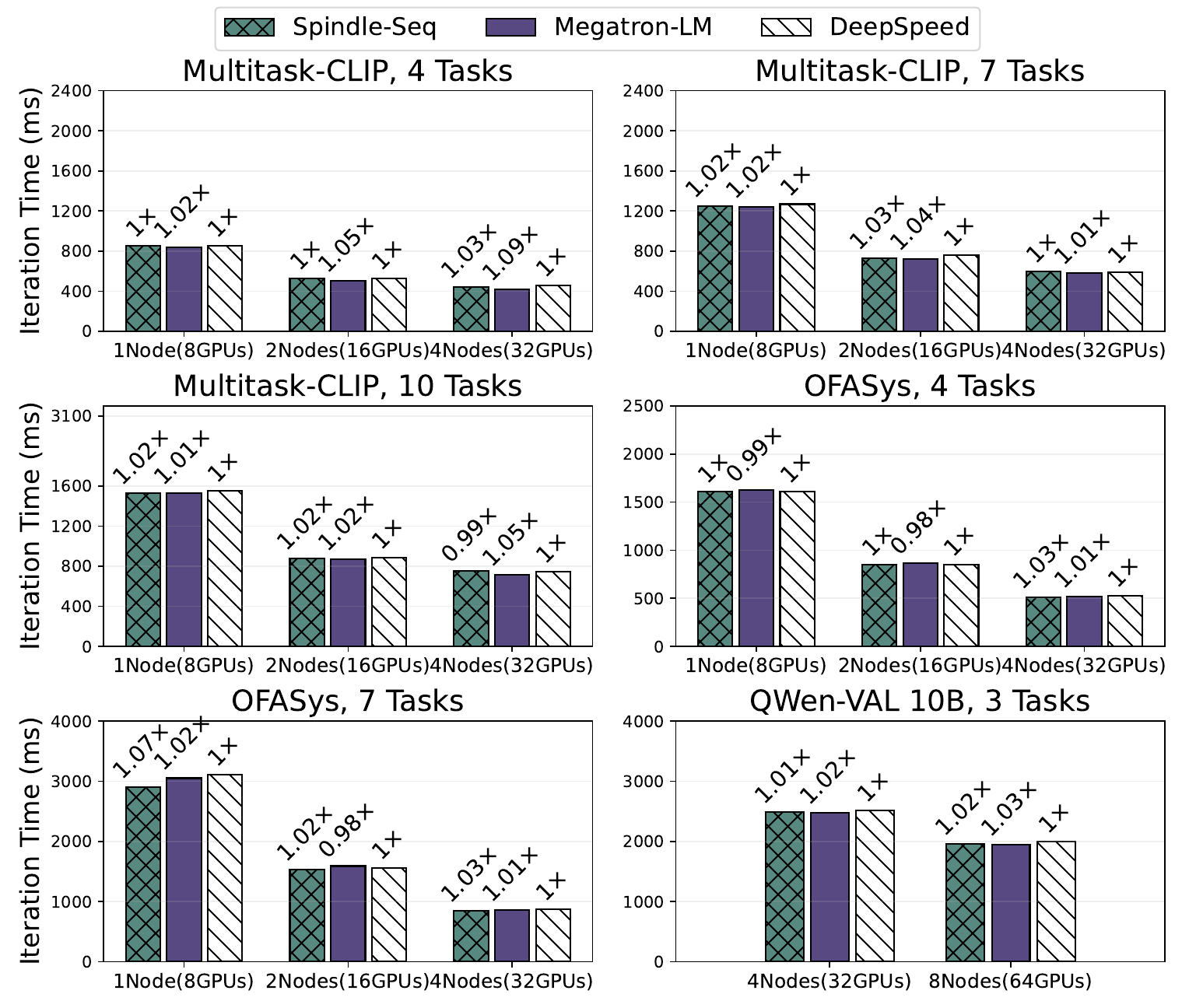}
    \caption{
        Comparison of \name-Seq with Megatron-LM and DeepSpeed.
    }
    \label{fig:seq_comparison}
\end{figure}

\fi

\end{document}